\documentclass[12pt]{article}
\usepackage{amssymb, amsmath, hyperref}
\usepackage{graphicx}
\usepackage{verbatim}
\def\ltwid{\mathrel{\raise.3ex\hbox{$<$\kern-.75em\lower1ex\hbox{$\sim$}}}}
\def\comp{{\rm C}\llap{\vrule height7.1pt width1pt depth-.4pt\phantom t}}
\def\square{\kern1pt\vbox{\hrule height 1.2pt\hbox{\vrule width 1.2pt\hskip 3pt
   \vbox{\vskip 6pt}\hskip 3pt\vrule width 0.6pt}\hrule height 0.6pt}\kern1pt}
\def\ltwid{\mathrel{\raise.3ex\hbox{$<$\kern-.75em\lower1ex\hbox{$\sim$}}}}

\begin{document}

\begin{titlepage}
\begin{flushright}
SPIN-09-14, ITP-UU-09/14, UFIFT-QG-09-01
\end{flushright}

\vspace{1cm}

\begin{center}
\bf{The Hubble Effective Potential}
\end{center}

\vspace{.3cm}

\begin{center}
T. M. Janssen$^{*}$, S. P. Miao$^{**}$, T. Prokopec$^{\ddagger}$
\end{center}
\begin{center}
\it{Institute for Theoretical Physics \& Spinoza Institute,
Utrecht University,\\
Leuvenlaan 4, Postbus 80.195, 3508 TD Utrecht, THE NETHERLANDS}
\end{center}

\vspace{0.2cm}

\begin{center}
R. P. Woodard$^{\dagger\dagger}$
\end{center}
\begin{center}
\it{Department of Physics, University of Florida \\
Gainesville, FL 32611, UNITED STATES.}
\end{center}

\vspace{.3cm}

\begin{center}
ABSTRACT
\end{center}
\hspace{0.3cm} We generalize the effective potential to scalar
field configurations which are proportional to the Hubble parameter
of a homogeneous and isotropic background geometry. This may be
useful in situations for which curvature effects are significant.
We evaluate the one loop contribution to the Hubble Effective Potential
for a massless scalar with arbitrary conformal and quartic couplings,
on a background for which the deceleration parameter is constant. Among 
other things, we find that inflationary particle production leads to 
symmetry restoration at late times.

\vspace{0.3cm}

\begin{flushleft}
PACS numbers: 04.30.-m, 04.62.+v, 98.80.Cq
\end{flushleft}

\vspace{.3cm}

\begin{flushleft}
$^{*}$ T.M.Janssen@uu.nl, \hspace{1cm} $^{**}$ S.Miao@uu.nl \\
$^{\ddagger}$ T.Prokopec@uu.nl, \hspace{1.2cm} $^{\dagger\dagger}$
woodard@phys.ufl.edu
\end{flushleft}

\end{titlepage}

\section{Introduction}

Consider two Heisenberg picture states, $\vert \Psi_1\rangle$ and $\vert \Psi_2
\rangle$, and a scalar field operator $\varphi(x)$ whose action is
$S[\varphi]$. We define the {\it background field} $\Phi(x)$ as the matrix
element of $\varphi(x)$ between these states,
\begin{equation}
\Bigl\langle \Psi_1 \Bigl\vert \varphi(x) \Bigr\vert \Psi_2\Bigr\rangle
\equiv \Phi(x) \; . \label{Phi}
\end{equation}
We define the {\it quantum field} $\phi(x)$ as the difference between
$\varphi(x)$ and $\Phi(x)$,
\begin{equation}
\varphi(x) \equiv \Phi(x) + \phi(x) \; . \label{phi}
\end{equation}
The ``effective field equation'' for $\Phi(x)$ is obtained by taking the
matrix element of the classical field equation,
\begin{equation}
\frac{\delta \Gamma[\Phi]}{\delta \Phi(x)} = \Biggl\langle \Psi_1 \Biggl\vert
\frac{\delta S[\varphi]}{\delta \varphi(x)} \Biggr\vert \Psi_2
\Biggr\rangle \; . \label{effeqn}
\end{equation}

These definitions might seem a little pointless because the Heisenberg field
$\varphi(x)$ must evolve so as to make $\delta S/\delta \varphi(x)$ vanish.
In that case (\ref{effeqn}) also vanishes and one would compute $\Phi(x)$ by
substituting the solution for $\varphi(x)$ into (\ref{Phi}). However, for
many purposes it is simpler to compute perturbative operator corrections to
$\phi(x)$ for arbitrary $\Phi(x)$, then substitute these into (\ref{effeqn})
and evaluate the matrix element assuming the matrix element for $\phi(x)$
vanishes. The result is a $\comp$-number functional of $\Phi(x)$ whose
vanishing determines the actual background field. Proceeding in this manner
is how one obtains the classic loop expansion for effective action
$\Gamma[\Phi]$ \cite{JS1,BSD1,LFA},
\begin{eqnarray}
\lefteqn{\Gamma[\Phi] = S[\Phi] + \frac{i \hbar}2 \ln\Biggl\{\det\Bigl[
\frac{\delta^2 S[\Phi]}{\delta \Phi \delta \Phi}\Bigr] \Biggr\} -
\frac{\hbar^2}8 \int \! d^D\!w \, d^D\!x \, d^D\!y \, d^D\!z} \nonumber \\
& & \hspace{.6cm} \times \frac{\delta^4 S[\Phi]}{\delta \Phi(w) \delta
\Phi(x) \delta \Phi(y) \delta \Phi(z)}
\Biggl[\frac{\delta^2 S[\Phi]}{\delta \Phi(w) \delta \Phi(x)}\Biggr]^{-1}
\Biggl[\frac{\delta^2 S[\Phi]}{\delta \Phi(y) \delta \Phi(z)}\Biggr]^{-1}
\nonumber \\
& & + \frac{\hbar^2}{12} \int \! d^D\!u \, d^D\!v \, d^D\!w
\, d^D\!x \, d^D\!y \, d^D\!z \, \frac{\delta^3 S[\Phi]}{\delta\Phi(u)
\delta\Phi(v) \delta \Phi(w)} \Biggl[\frac{\delta^2 S[\Phi]}{\delta
\Phi(u) \delta \Phi(x)}\Biggr]^{-1} \qquad \nonumber \\
& & \hspace{.6cm} \times
\Biggl[\frac{\delta^2 S[\Phi]}{\delta \Phi(v) \delta \Phi(y)}\Biggr]^{-1}
\Biggl[\frac{\delta^2 S[\Phi]}{\delta \Phi(w) \delta \Phi(z)}\Biggr]^{-1}
\frac{\delta^3 S[\Phi]}{\delta\Phi(x) \delta\Phi(y) \delta \Phi(z)}
+ O(\hbar^3) \; . \qquad
\end{eqnarray}
Here the exponent $\mbox{}^{-1}$ denotes the functional inverse,
\begin{equation}
\int \! d^D\!y \, \frac{\delta^2 S[\Phi]}{\delta \Phi(x) \delta \Phi(y)}
\Biggl[\frac{\delta^2 S[\Phi]}{\delta \Phi(y) \delta \Phi(z)}\Biggr]^{-1}
= \delta^D(x \!-\! z) \; . \label{inverse}
\end{equation}

The scheme described above obviously depends a little upon the states 
$\vert \Psi_i \rangle$. For the familiar, in-out effective field equations 
the state $\vert \Psi_2 \rangle$ is free vacuum (centered around $\Phi$) in 
the infinite past, while $\vert \Psi_1\rangle$ is free vacuum (again centered
on $\Phi$) in the infinite future. For the Schwinger-Keldysh effective
field equations one has $\vert \Psi_1\rangle = \vert \Psi_2 \rangle$, and 
they are taken to be free vacuum (centered on $\Phi$) in the infinite past
\cite{JS2,others}. One can consider variants of the Schwinger-Keldysh 
equations in which the two states are still the same but the two states are 
defined at some finite time \cite{Berges:2004yj,FW}. This is especially 
convenient for cosmological settings in which the initial singularity 
precludes reaching the infinite past, and we do not know what is a reasonable 
state to assume for the infinite future.

The hard part about computing the effective action to a fixed loop order
is solving for the functional inverses (\ref{inverse}). Multiplying by
$i$ (times $\hbar$ which we henceforth set to one) gives the propagator
in the presence of background $\Phi$,
\begin{equation}
i \Biggl[\frac{\delta^2 S[\Phi]}{\delta \Phi(x) \delta \Phi(x')}\Biggr]^{-1}
\equiv i\Delta[\Phi](x;x') \; .
\end{equation}
An elegant formalism exists for deriving asymptotic expansions which are
useful near coincidence, i.e., for $x^{\prime \mu}$ near $x^{\mu}$
\cite{BSD2,BSD1,RTS}. This allows one to give a wonderfully general 
treatment of ultraviolet divergences \cite{BSD3,SMC,BV}, but it sacrifices 
the nonlocal, ultraviolet finite contributions which constitute some of the 
most potentially interesting quantum effects.

Solving for this propagator is typically so difficult that one is forced to
specialize the effective action to a handful of special backgrounds. One of
these special backgrounds, and one that plays an important role in flat 
space physics in trivial backgrounds, is $\Phi(x) = \Phi_0$, where $\Phi_0$ 
is a constant in space and time. In that case the dependence of the kinetic 
operator on $\Phi_0$ typically reduces to a mass term and we can compute 
the propagator.

The background $\Phi(x) = \Phi_0$ defines the {\it effective potential},
\begin{equation}
\Gamma[\Phi_0] \equiv -\int \! d^D\!x \times V_{\rm eff}(\Phi_0) \; .
\end{equation}
In some cases $V_{\rm eff}(\Phi_0)$ is precisely what we want. For example,
if there is a vacuum state in flat space quantum field theory then it must
be invariant under spacetime translations, so the actual solution is 
constant and we are guaranteed to find it by minimizing the effective 
potential. The value of the effective potential at this solution gives the 
vacuum energy density, and the second derivative of $V_{\rm eff}$ at the 
solution determines the scalar mass.

However, the more usual situation is that we would like to have the full 
effective action, and we only settle for the effective potential because 
it is what can be computed. There is often no point in working out the
effective action for a constant background field versus some other 
configuration with spacetime dependence. It is especially pointless to 
demand that the background be constant in cosmological settings, for which 
the geometry and many of the fields evolve in time.

In this paper we shall explore a different class of backgrounds, namely
those for which $\Phi(x)$ is proportional to the Hubble parameter $H(t)$,
\begin{equation}
\Phi(t,\vec{x}) = \Phi_0 \times H(t) \; . \label{hubbak}
\end{equation}
Of course the classical part of this {\it Hubble Effective Potential} ---
let us call it $V_{\rm hub}(\Phi_0)$ --- is straightforward to read off from 
the classical action. The derivative of the quantum correction to 
$V_{\rm hub}$ is defined by the difference between the full effective field 
equations and the classical ones at the background (\ref{hubbak}),
\begin{equation}
\Biggl\{ \frac{\delta \Gamma[\Phi]}{\delta \Phi(x)} -
\frac{\delta S[\Phi]}{\delta \Phi(x)} \Biggr\}_{\Phi = \Phi_0 H} \equiv
-\Delta V_{\rm hub}'(\Phi_0 H) \sqrt{-g} \; .
\end{equation}
As is apparent from the possibility of a conformal coupling in the 
classical Lagrangian, the dependence of $V_{\rm hub}(\Phi_0 H)$ on
$H(t)$ is not limited to its argument $\Phi_0 H$. We will see that 
quantum corrections to $V_{\rm hub}$ can also depend explicitly upon time.

Knowing the Hubble Effective Potential might be useful in situations for 
which the scalar tends to evolve in tandem with the expansion of the 
universe. To see that this can occur in the early universe it suffices to 
study the classical equation of motion for a spatially homogeneous, 
quarticly coupled scalar on a homogeneous and isotropic background,
\begin{equation}
\ddot{\Phi}(t) + 3 H(t) \, \dot{\Phi}(t) + \Bigl[ m^2 + \xi R(t)\Bigr] \Phi(t)
+ \frac{\lambda}6 \, \Phi^3(t) = 0 \; , \label{class}
\end{equation}
where $m$ is the scalar mass, $\xi$ the conformal coupling, $\lambda$
is the quartic coupling constant and $R = 6\dot{H} + 12 H^2$ is the Ricci 
scalar. In the limit of large $H$ the mass term becomes irrelevant. Let
us also assume that the deceleration is constant, which is nearly the case 
during long epochs of cosmological history. In that case we have,
\begin{equation}
\dot{H} = -\epsilon H^2 \qquad {\rm and} \qquad \ddot{H} = 2\epsilon^2 H^3 \; ,
\end{equation}
for some constant $\epsilon$. Under these assumptions --- which certainly
pertain at early enough times --- substituting the Hubble scaling ansatz 
(\ref{hubbak}) into (\ref{class}) gives the following cubic equation for 
the proportionality factor $\Phi_0$,
\begin{equation}
\Bigl[2 \epsilon^2 - 3\epsilon + 6 (2 - \epsilon) \xi \Bigr] \Phi_0 +
\frac{\lambda}6 \Phi_0^3 = 0 \; .
\end{equation}
It is not difficult to imagine values of $\xi$ and $\epsilon$ (e.g., 
power law inflation) for which the two symmetry-breaking solutions are 
real and the field tracks the evolution of the Hubble parameter. At
least that would be the classical result; the Hubble Effective Potential 
describes how quantum corrections might change this.

The usual Effective Potential $V_{\rm eff}(\Phi_0)$ tells us the energy 
density of the minimum uncertainty state centered on $\Phi_0$. What the
Hubble Effective Potential gives us is that part of the energy density
of the minimum uncertainty state centered on $\Phi_0 H(t)$ which is not
attributable to the kinetic energy of the scalar background. Of course
this includes the usual result from $V_{\rm eff}$ but it also receives
contributions from the way the expansion of spacetime excites the scalar.
When spacetime expansion leads to physical particle production there can
be secular effects.

This paper consists of five sections, of which the first is nearing its end.
In section 2 we lay out the Feynman rules for using dimensional regularization
to compute in massless $\lambda \varphi^4$ theory on a homogeneous and 
isotropic background geometry of constant deceleration. These rules are 
employed in section 3 to obtain a fully renormalized result for the one 
loop correction to the Hubble Effective Potential. One fascinating feature 
of the result is secular dependence in addition to that inherited from the 
simple fact that the scalar background is being evaluated at a time-dependent 
value (\ref{hubbak}). This is discussed in section 4, as is the dependence 
upon the arbitrary (but constant) value of the deceleration parameter. Our 
conclusions comprise section 5.

\section{Feynman Rules}

The purpose of this section is to give the Feynman rules for computing
loop corrections for a massless scalar with a quartic self-interaction
on a nondynamical background geometry with constant deceleration. We
begin with a discussion of the geometry, then present the bare Lagrangian
and the counterterms. The section closes with a discussion of the 
propagator in a general Hubble-scaled background (\ref{hubbak}). This is
the key result which has made the present work possible.

On the largest scales, the visible universe is well described by a
homogeneous, isotropic and spatially flat geometry. The invariant
element takes a familiar form when expressed either in co-moving 
$(t,\vec{x})$ or conformal $(\eta,\vec{x})$ coordinates,
\begin{equation}
ds^2 \equiv g_{\mu\nu} dx^{\mu} dx^{\nu} = -dt^2 + a^2(t) d\vec{x}
\!\cdot\! d\vec{x} = a^2 \Bigl[-d\eta^2 + d\vec{x} \!\cdot\! d\vec{x}\Bigr]
\; . \label{geom}
\end{equation}
To facilitate dimensional regularization we shall work in $D$ spacetime
dimensions, so the indices $\mu$ and $\nu$ run from $0$ to $(D\!-\!1)$.

Two derivatives of the scale factor $a(t)$ have great importance,
the Hubble parameter $H(t)$ and the deceleration parameter $q(t)$,
\begin{equation}
H(t) \equiv \frac{\dot{a}}{a} \qquad , \qquad q(t) \equiv -1 -
\frac{\dot{H}}{H^2} \equiv -1 + \epsilon(t) \; . \label{Hq}
\end{equation}
Although the Hubble parameter may have changed by as much as 57
orders of magnitude from primordial inflation to now, the
deceleration parameter is only thought to have varied from nearly
$-1$, during primordial inflation, to $+1$, during the epoch of 
radiation domination. Because $q(t)$ has been approximately
constant for vast periods of cosmological evolution it is
interesting to consider physics during an epoch of constant 
deceleration. 

It is simple to reconstruct the Hubble parameter and the scale factor
over any epoch during which $\epsilon$ is constant. If we define the
initial values as $H(0) \equiv H_0$ and $a(0) \equiv 1$ then relation 
(\ref{Hq}) implies,
\begin{equation}
H(t) = \frac{H_0}{1 \!+\! \epsilon H_0 t} \qquad {\rm and} \qquad a(t) 
= [1 \!+\! \epsilon H_0 t]^{\frac1{\epsilon}} \; .
\end{equation}
The conformal time $\eta$ is defined by $d\eta = dt/a(t)$ up to a constant
of integration. We can choose this constant to give equivalent expressions 
for $H$ and $a$,
\begin{equation}
H = \frac{H_0}{[-(1\!-\!\epsilon) H_0 \eta]^{\frac{-\epsilon}{1-\epsilon}}}
\qquad {\rm and} \qquad a = \frac1{[-(1\!-\!\epsilon) H_0 \eta]^{\frac1{
1-\epsilon}}} \; .
\end{equation}
Note that the universe is accelerating for $0 \leq \epsilon < 1$, and our
conventions cause $\eta$ to approach zero fom below at late times. For
$\epsilon > 1$ the universe is decelerating and $\eta$ is positive. Note
also the simple and very useful relation,
\begin{equation}
(1\!-\!\epsilon) H a = -\frac1{\eta} \; . \label{useful}
\end{equation}
Finally, it is worth noting that the restriction to constant $\epsilon$
reduces all curvature invariants to algebraic functions of $\epsilon$ times 
powers of $H$,
\begin{eqnarray}
R & = & (D\!-\!1) (D \!-\! 2 \epsilon) H^2 \; , \\
\square R & = & 2 (D\!-\!1) \epsilon (D \!-\!2\epsilon) (D\!-\!1 \!-\! 3 
\epsilon) H^4 \; , \\
R^{\mu\nu} R_{\mu\nu} & = & (D\!-\!1) \Bigl[(D\!-\!1)D - 4(D\!-\!1) \epsilon
+ D \epsilon^2\Bigr] H^4 \; , \\
R^{\rho\sigma\mu\nu} R_{\rho\sigma\mu\nu} & = & 2 (D\!-\!1) \Bigl[D - 
4\epsilon + 2 \epsilon^2\Bigr] H^4 \; .
\end{eqnarray}

The bare Lagrangian of interest for us is,
\begin{equation}
\mathcal{L} = -\frac12 \partial_{\mu} \varphi_0 \partial_{\nu} \varphi_0
g^{\mu\nu} \sqrt{-g} -\frac12 \xi_0 \varphi_0^2 R \sqrt{-g} -\frac1{4!}
\lambda_0 \varphi_0^4 \sqrt{-g} \; . \label{bare}
\end{equation}
Here $\varphi_0$ is the bare field and $\xi_0$ and $\lambda_0$ are the
bare conformal and quartic coupling constants. We assume zero bare mass and,
because mass is multiplicatively renormalized in dimensional regularization,
we shall not require a mass counterterm. 

The renormalized field $\varphi$ is defined in the usual way,
\begin{equation}
\varphi \equiv Z^{-\frac12} \varphi_0 \; .
\end{equation}
This gives the following expression for the bare Lagrangian (\ref{bare}),
\begin{equation}
\mathcal{L} = -\frac12 Z \partial_{\mu} \varphi \partial_{\nu} \varphi
g^{\mu\nu} \sqrt{-g} -\frac12 Z \xi_0 \varphi^2 R \sqrt{-g} -\frac1{4!}
Z^2 \lambda_0 \varphi^4 \sqrt{-g} \; . \label{L2}
\end{equation}
Renormalization is implemented by defining the bare couplings in terms of 
the physical ones plus counter parameters,
\begin{equation}
Z \equiv 1 + \delta Z \qquad , \qquad Z \xi_0 \equiv \xi + \delta \xi
\qquad {\rm and} \qquad Z^2 \lambda_0 \equiv \lambda + \delta \lambda \; .
\end{equation}
One then expresses the Lagrangian in terms of primitive interactions 
involving the renormalized fields and couplings, plus a series of
counterterms,
\begin{eqnarray}
\lefteqn{\mathcal{L} = -\frac12 \partial_{\mu} \varphi \partial_{\nu} \varphi
g^{\mu\nu} \sqrt{-g} -\frac12 \xi \varphi^2 R \sqrt{-g} -\frac1{4!}
\lambda \varphi^4 \sqrt{-g} } \nonumber \\
& & \hspace{2cm} -\frac12 \delta Z \partial_{\mu} \varphi \partial_{\nu} 
\varphi g^{\mu\nu} \sqrt{-g} -\frac12 \delta \xi \varphi^2 R \sqrt{-g} 
-\frac1{4!} \delta \lambda \varphi^4 \sqrt{-g} \; . \qquad \label{L3}
\end{eqnarray}
We consider $\xi$ to be a free parameter of order one in $\lambda$. The 
various counter parameters are of order,
\begin{equation}
\delta Z = O(\lambda^2) \qquad , \qquad \delta \xi = O(\lambda) \qquad
{\rm and} \qquad \delta \lambda = O(\lambda^2) \; .
\end{equation}
Therefore $\delta Z$ is not required at one loop while both $\delta \xi$
and $\delta \lambda$ are.

Of course the formalism we have just reviewed is valid for any background
geometry and any scalar field. If only one could do the same for the
propagator! The equation it obeys is simple enough to write down for 
arbitrary metric and scalar backgrounds,
\begin{equation}
\Bigl[\partial_{\mu} \sqrt{-g} g^{\mu\nu} \partial_{\nu}
-\xi R \sqrt{-g} - \frac12 \lambda \Phi^2 \sqrt{-g} \Bigr] 
i\Delta(x;x') = i \delta^D(x-x') \; . \label{propeq}
\end{equation}
Alas, the set of known solutions is restricted to a pitiful handful of
backgrounds. 

The recent technical advance which enables the present work is a tractable
resolution for the problem of infrared divergences in the closely related 
propagator for the massless, minimally coupled scalar on a background of 
arbitrary, constant deceleration \cite{us}. To make the connection, let us
use conformal coordinates, and some facts about our constant deceleration
geometry, to re-express the massless, minimally coupled kinetic operator,
\begin{eqnarray}
\partial_{\mu} \sqrt{-g} g^{\mu\nu} \partial_{\nu} & = & \eta^{\mu\nu}
\partial_{\mu} a^{D-2} \partial_{\nu} \; , \\
& = & a^{\frac{D}2-1} \Bigl\{ \eta^{\mu\nu} \partial_{\mu} \partial_{\nu}
+ \frac14 \Bigl(\frac{D\!-\!2}{D\!-\!1}\Bigr) \, R a^2 \Bigr\} 
a^{\frac{D}2-1} \; , \\
& = & a^{\frac{D}2-1} \Bigl\{ \eta^{\mu\nu} \partial_{\mu} \partial_{\nu}
+ \frac{(D\!-\!2) (D \!-\! 2 \epsilon)}{4 (1\!-\! \epsilon)^2 \eta^2}
\Bigr\} a^{\frac{D}2-1} \; , \\
& = & a^{\frac{D}2-1} \Bigl\{ \eta^{\mu\nu} \partial_{\mu} \partial_{\nu}
+ \frac{\nu^2_{\rm mmc} \!-\! \frac14}{\eta^2} \Bigr\} a^{\frac{D}2-1} \; .
\end{eqnarray}
Here the index $\nu_{\rm mmc}$ is,
\begin{equation}
\nu_{\rm mmc} = \frac12 \Bigl(\frac{D \!-\! 1 \!-\! \epsilon}{1 \!-\!
\epsilon}\Bigr) \; . \label{nummc}
\end{equation}
Multiplying by the factor $(a'/a)^{\frac{D}2-1}$ --- where $a$ is the scale
factor at $x^{\mu}$ and $a'$ is the scale factor at $x^{\prime \mu}$ ---
therefore carries the equation for the propagator of a massless, minimally
coupled scalar,
\begin{equation}
\partial_{\mu} \sqrt{-g} g^{\mu\nu} \partial_{\nu} \, i\Delta_{\rm mmc}(x;x')
= i \delta^D(x \!-\! x') \; ,
\end{equation}
to the form,
\begin{equation}
\Bigl[ \eta^{\mu\nu} \partial_{\mu} \partial_{\nu} + \frac{\nu_{\rm mmc}^2 
\!-\! \frac14}{\eta^2} \Bigr] \Bigl\{ (a a')^{\frac{D}2-1} i
\Delta_{\rm mmc}(x;x')\Bigr\} = i \delta^D(x \!-\! x') \; . \label{mmceq}
\end{equation}

A point of great importance for us is that equation (\ref{mmceq}), with
the replacement $\nu_{\rm mmc} \rightarrow \nu$, can be solved {\it for any} 
constant index $\nu$, whether or not $\nu$ happens to obey (\ref{nummc}) 
\cite{Bunch:1977sq,JP1,JMP}. Now note that, for a scalar background 
$\Phi(t,\vec{x}) = \Phi_0 H(t)$, in a geometry of constant deceleration, 
the kinetic operator in (\ref{propeq}) can be reduced to precisely this
form,
\begin{eqnarray}
\lefteqn{\partial_{\mu} \sqrt{-g} g^{\mu\nu} \partial_{\nu}
-\xi R \sqrt{-g} - \frac12 \lambda \Phi^2 \sqrt{-g} } \nonumber \\
& & \hspace{2cm} = \eta^{\mu\nu} \partial_{\mu} a^{D-2} \partial_{\nu} - 
\xi (D\!-\!1) (D \!-\!2\epsilon) H^2 a^D - \frac12 \lambda \Phi_0^2 H^2 a^D 
\; , \qquad \\
& & \hspace{2cm} = a^{\frac{D}2-1} \Bigl\{ \eta^{\mu\nu} \partial_{\mu} 
\partial_{\nu} + \frac{\nu^2 \!-\! \frac14}{\eta^2} \Bigr\} a^{\frac{D}2-1} \;,
\end{eqnarray}
where the index $\nu$ obeys,
\begin{equation}
\nu^2 = \frac14 \Bigl(\frac{D \!-\! 1 \!-\! \epsilon}{1 \!-\! \epsilon}\Bigr)^2
- \frac{\xi (D\!-\!1) (D\!-\!2\epsilon) + \frac12 \lambda \Phi_0^2}{(1 \!-\!
\epsilon)^2} \; . \label{ournu}
\end{equation}
Hence we need only adapt the index of the previous solution
\cite{Bunch:1977sq,JP1,JMP} once the infrared problem is solved.

A formal solution to equation (\ref{mmceq}) was found many years ago by 
Bunch and Davies for the spacetime manifold $R^4$ \cite{Bunch:1977sq}.
The generalization of this infinite space solution to $D$ dimensions is 
\cite{JP1,JMP},
\begin{eqnarray}
\lefteqn{i\Delta_{\infty}(x;x') = \frac{[(1\!-\!\epsilon)^2 H H']^{
\frac{D}2-1}}{(4\pi)^{\frac{D}2}} \, \frac{\Gamma(\frac{D-1}2 \!+\!\nu) 
\Gamma(\frac{D-1}2\!-\! \nu)}{\Gamma(\frac{D}2)} } \nonumber \\
& & \hspace{5cm} \times \; \mbox{}_2 F_1\Bigl(\frac{D\!-\!1}2\!+\!
\nu,\frac{D\!-\!1}2\!-\!\nu;\frac{D}2;1\!-\!\frac{y}4\Bigr) \; , \qquad
\label{Tomislav}
\end{eqnarray}
where $H$ and $H'$ are the Hubble parameters at $x^{\mu}$ and $x^{\prime \mu}$,
respectively, and the length function $y(x;x')$ is,\footnote{Note the 
unfortunate similarity between the geometrical parameter, $\epsilon \equiv 
-\dot{H}/H^2$, and the infinitesimal parameter $\varepsilon$ which enforces 
Feynman boundary conditions.}
\begin{equation}
y(x;x') \equiv \frac{\Vert \vec{x} \!-\! \vec{x}'\Vert^2 - (\vert \eta \!-\!
\eta'\vert - i \varepsilon)^2}{\eta \eta'} \; .
\end{equation}
As stated, the solution appropriate for the massless, minimally coupled
scalar has the index $\nu = \nu_{\rm mmc}$, but (\ref{Tomislav}) solves
our propagator equation (\ref{propeq}) as well, with the index $\nu$ 
set to (\ref{ournu}).

The infinite space solution (\ref{Tomislav}) is only formal because it
possesses infrared divergences in $D=4$ \cite{FP} which the $D$-dimensional 
version inherits \cite{JP2}. The presence of an infrared divergence in
quantum field theory signals that the quantity which diverges is unphysical 
in some way. In this case the problem is that one is assuming the initial 
state could have been prepared in coherent Bunch-Davies vacuum out to 
infinite wavelengths. No causal process can accomplish this, so it is 
neither surprising nor problematic that making the assumption results in an 
infrared divergence. The assumption can be relaxed in either of two ways:
\begin{enumerate}
\item{One might continue to assume that the spatial sections are $R^{D-1}$
but that the initially super-horizon wavelengths are in some less singular
state \cite{AV}. That will guarantee there are no initial infrared 
singularities, which suffices to show that none develop later in time
\cite{Fulling:1978ht}.}
\item{One might instead work on the spatial manifold $T^{D-1}$, which means
there are no super-horizon wavelengths past a certain value and hence no
possibility for an infrared divergence \cite{TW3}.}
\end{enumerate}

In \cite{us} we obtained an explicit result for the latter fix. This makes
the mode sum discrete, however, one can make the integral approximation with
the lower limit of the co-moving momentum cutoff at some fixed value $k_0$.
Then the corrections to the infinite space result (\ref{Tomislav}) consist of
the same integrand, integrated from $\Vert \vec{k}\Vert = 0$ to $\Vert \vec{k}
\Vert = k_0$. Most of these corrections actually vanish for $k_0 = 0$, and it
is only necessary to include the ones which grow. To find them, note from
(\ref{Tomislav}) that the infinite space result diverges whenever the index 
$\nu$ in equation (\ref{ournu}) obeys the relation,
\begin{equation}
\vert \nu \vert \longrightarrow \nu_N \equiv  \Bigl(\frac{D\!-\!1}2\Bigr) 
+ N \; ,
\end{equation}
for some nonnegative integer $N$. These correspond to the special values of 
$\nu$ for which there is a logarithmic infrared divergence that dimensional 
regularization registers. The lower limit term which subtracts off this 
divergence is \cite{us},
\begin{eqnarray}
\lefteqn{\delta i\Delta_N(x;x') = \frac{[(1-\epsilon)^2H H']^{\frac{D}{2}-1}}{
(4\pi)^{\frac{D}{2}}} \, \frac{\Gamma(2\nu) \Gamma(\nu)}{\Gamma(\frac12 \!+\!
\nu) \Gamma(\frac{D-1}2)} \, \frac1{\nu \!-\! \nu_N} } \nonumber \\
& & \hspace{2.2cm} \times \Bigl(k_0^2 \eta \eta'\Bigr)^{\frac{D-1}2 -\nu} \,
k_0^{2N} \sum_{k=0}^N \sum_{\ell=0}^{N-k} \alpha_{k\ell} \Vert \vec{x} \!-\! 
\vec{x}' \Vert^{2k} \eta^{2\ell} \eta'^{2(N-k-\ell)} \; , \qquad \label{idD}
\end{eqnarray}
with the coefficients $\alpha_{k\ell}$ defined as,
\begin{equation}
\alpha_{k\ell} = \frac{(-\frac14)^N}{k! \, \ell! \, (N \!-\! k \!-\! \ell)!} 
\, \frac{\Gamma(\frac{D-1}2) \, \Gamma^2(1 \!-\! \nu)}{\Gamma(k \!+\! 
\frac{D-1}2) \Gamma(\ell \!+\! 1 \!-\! \nu) \Gamma(N \!-\! k \!-\! \ell \!+\! 
1 \!-\! \nu)} \; . \label{akl}
\end{equation}

For $N > 0$ the infinite space mode sum also contains power law infrared 
divergences that dimensional regularization incorrectly removes by automatic 
subtraction \cite{us}. These are removed by lower limit corrections which
correspond to $\delta i\Delta_M(x;x')$ for $0 \leq M< N$. All the other
corrections vanish for $k_0 = 0$, so the full propagator for $\nu$ near 
$\nu_N$ is,
\begin{equation}
i\Delta(x;x') \approx i\Delta_{\infty}(x;x')
+ \sum_{M=0}^N \delta i\Delta_M(x;x') \; . \label{fullprop}
\end{equation}
We have shown \cite{us} that this form agrees with results previously 
obtained for the special cases of $\epsilon = 0$ for general $D$ \cite{OW},
and $\epsilon = \frac12$ for $D =4$ \cite{ITTW}.

\section{Hubble Effective Potential at One Loop}

The purpose of this section is to compute and renormalize the one 
loop correction to effective field equation of massless $\varphi^4$
theory for an arbitrary geometry of constant deceleration in the
Hubble-scaled background (\ref{hubbak}),
\begin{equation}
\Phi(t,\vec{x}) = \Phi_0 \times H(t) \; .
\end{equation}
We begin by writing the one loop effective field equations in general,
then specialize to constant deceleration and the Hubble-scaled scalar
background. This permits us to evaluate the coincident scalar propagator
using our results (\ref{Tomislav}) and (\ref{idD}-\ref{fullprop}).

The Schwinger-Keldysh effective field equation for $\Phi(x)$ is obtained 
by taking the expectation value of the Heisenberg operator equation
for $\varphi(x)$,
\begin{eqnarray}
\lefteqn{\frac{\delta S[\varphi]}{\delta \varphi(x)} = 
\partial_{\mu} \Bigl(\sqrt{-g} g^{\mu\nu} \partial_{\nu} \varphi\Bigr) 
- \xi \varphi R \sqrt{-g} - \frac{\lambda}{6} \, \varphi^3 \sqrt{-g} } 
\nonumber \\
& & \hspace{3cm} + \delta Z \partial_{\mu} \Bigl(\sqrt{-g} g^{\mu\nu} 
\partial_{\nu} \varphi\Bigr) - \delta \xi \varphi R \sqrt{-g} 
-\frac{\delta \lambda}{6} \, \varphi^3 \sqrt{-g} \; . \qquad
\end{eqnarray}
One breaks the full field $\varphi(x)$ up into its $\comp$-number 
expectation value $\Phi(x)$ and the quantum field $\phi(x)$,
\begin{equation}
\varphi(x) = \Phi(x) + \phi(x) \qquad {\rm where} \qquad \langle \Omega 
\vert \phi(x) \vert \Omega \rangle = 0 \; .
\end{equation}
At one loop the result is,
\begin{eqnarray}
\lefteqn{\Bigl\langle \Omega \Bigl\vert \frac{\delta S[\varphi]}{\delta 
\varphi(x)} \Bigr\vert \Omega \Bigr\rangle = \partial_{\mu} \Bigl(\sqrt{-g} 
g^{\mu\nu} \partial_{\nu} \Phi\Bigr) - \xi \Phi R \sqrt{-g} - 
\frac{\lambda}{6} \, \Phi^3 \sqrt{-g} } \nonumber \\
& & \hspace{1.4cm} - \frac{\lambda}{2} \, \Phi \langle \Omega \vert \phi^2(x) 
\vert \Omega \rangle \sqrt{-g} - \delta \xi \Phi R \sqrt{-g} -\frac{\delta 
\lambda}{6} \, \Phi^3 \sqrt{-g} + O(\hbar^2) \; . \qquad \label{fullone}
\end{eqnarray}

The effective field equation (\ref{fullone}) is valid for any metric and 
scalar background, but it is of course impossible to obtain explicit results 
for the coincident scalar propagator which is implicit in $\langle \Omega 
\vert \phi^2(x) \vert \Omega \rangle$, except in special backgrounds. We 
therefore specialize at this point to the geometry of arbitrary but constant 
deceleration described in the previous section, and to the Hubble-scaled 
scalar (\ref{hubbak}),
\begin{eqnarray}
\langle \Omega \vert \phi^2(x) \vert \Omega \rangle & = & i\Delta(x;x) +
O(\hbar^2) \; , \\
& = & i\Delta_{\infty}(x;x) + \sum_{M=0}^N \delta i\Delta_M(x;x) +
O(\hbar^2) \; . \label{2terms}
\end{eqnarray}

Of the various one loop contributions to (\ref{2terms}), only the infinite 
space propagator harbors ultraviolet divergences, so we will evaluate it,
renormalize and then include the finite contributions from $\delta i
\Delta_{M}(x;x)$. The coincidence limit of the infinite space propagator 
(\ref{Tomislav}) is,
\begin{equation}
i\Delta_{\infty}(x;x) = \frac{[(1\!-\!\epsilon)^2 H^2]^{\frac{D}2-1}}{
(4\pi)^{\frac{D}2}} \, \frac{\Gamma(\frac{D-1}2 \!+\! \nu) \Gamma(\frac{D-1}2
\!-\! \nu) \Gamma(1 \!-\! \frac{D}2)}{\Gamma(\frac12 \!+\! \nu) 
\Gamma(\frac12 \!-\! \nu)} \; . \label{diverge}
\end{equation}
At this point we define $\delta \equiv 4 - D$ and expand the $\nu$-dependent
ratios,
\begin{eqnarray}
\lefteqn{ \frac{\Gamma(\frac{D-1}2 \!+\! \nu) \Gamma(\frac{D-1}2 \!-\! \nu)}{
\Gamma(\frac12 \!+\! \nu) \Gamma(\frac12 \!-\! \nu)} =
\Bigl[ \Bigl(\frac{D\!-\!3}2\Bigr)^2 - \nu^2\Bigr] \,
\frac{\Gamma(\frac{D-3}2 \!+\! \nu) \Gamma(\frac{D-3}2 \!-\! \nu)}{
\Gamma(\frac12 \!+\! \nu) \Gamma(\frac12 \!-\! \nu)} \; , } \\
& & \hspace{-.5cm} = \frac1{(1 \!-\! \epsilon)^2} \, \Biggl\{ \Bigl(\xi \!-\! 
\frac16\Bigr) (D\!-\!1) (D\!-\! 2\epsilon) + \frac{\lambda}2 \, \Phi_0^2 - 
\frac16 \Bigl(1 \!-\! 5\epsilon \!+\! 3 \epsilon^2\Bigr) \delta + 
O(\delta^2)\Biggr\} \nonumber \\
& & \hspace{4cm} \times \Biggl\{1 - \Bigl[ \psi\Bigl(\frac12 \!+\! \nu\Bigr) + 
\psi\Bigl(\frac12 \!-\! \nu\Bigr) \Bigr] \frac{\delta}2 + O(\delta^2)
\Biggr\} \; . \qquad \label{exp1}
\end{eqnarray}
The expansion of the divergent Gamma function is also important,
\begin{equation}
\Gamma\Bigl(1 \!-\! \frac{D}2\Bigr) = -\frac2{\delta} - 1 + \gamma 
+ O(\delta) \; , \label{exp2}
\end{equation}
where $\gamma \approx .577216$ is the Euler constant.
Substituting (\ref{exp1}-\ref{exp2}) in (\ref{diverge}) gives,
\begin{eqnarray}
\lefteqn{ i\Delta_{\infty}(x;x) = \frac{\Gamma(1 \!-\! \frac{D}2)}{
(4\pi)^{\frac{D}2} [(1\!-\!\epsilon)^2 H^2]^{\frac{\delta}2}}
\Bigl[ \Bigl(\xi \!-\! \frac16\Bigr) R + \frac{\lambda}2 \, \Phi^2\Bigr] } 
\nonumber \\
& & \hspace{2cm} + \frac1{16 \pi^2} \Bigl[ \Bigl(\xi \!-\! \frac16\Bigr) R
+ \frac{\lambda}2 \, \Phi^2\Bigr] \Bigl[\psi\Bigl( \frac12 \!+\! \nu) +
\psi\Bigl(\frac12 \!-\! \nu\Bigr)\Bigr] \nonumber \\
& & \hspace{6cm} + \frac1{16 \pi^2} \, \frac13 (1 \!-\! 5\epsilon \!+\! 
3 \epsilon^2) H^2 + O(\delta) \; . \qquad \label{divexp}
\end{eqnarray}

Employing (\ref{2terms}) and (\ref{divexp}) in the effective field equation 
(\ref{fullone}) shows that we must make the following choices for the one
loop counter parameters,
\begin{eqnarray}
\delta \xi & = & -\frac{\lambda}2 \, \Bigl(\xi \!-\! \frac16\Bigr) \,
\frac{\Gamma(1 \!-\! \frac{D}2)}{(4 \pi)^{\frac{D}2} \mu_1^{\delta}} +
O(\lambda^2) \; , \\
\delta \lambda & = & -\frac{3 \lambda^2}2 \, \frac{\Gamma(1 \!-\! \frac{D}2)}{
(4 \pi)^{\frac{D}2} \mu_2^{\delta}} + O(\lambda^3) \; .
\end{eqnarray}
Here $\mu_1$ and $\mu_2$ are two dimensional regularization mass scales which
can be used to subsume the finite parts of the counter parameters. With these
choices we can take the unregulated limit to obtain $V_{\rm hub}'(\Phi)$,
\begin{eqnarray}
\lefteqn{V_{\rm hub}'(\Phi) = 
\Biggl\{ \xi + \frac{\lambda (\xi \!-\! \frac16)}{32 \pi^2} \Biggl[
\ln\Bigl[ \frac{(1 \!-\! \epsilon)^2 H^2}{\mu_1^2}\Bigr] + \psi\Bigl(\frac12
\!+\! \nu\Bigr) + \psi\Bigl(\frac12 \!-\! \nu\Bigr)\Biggr]\Biggr\} R \Phi }
\nonumber \\
& & \hspace{1.5cm} + \Biggl\{ \frac{\lambda}6 + \frac{\lambda^2}{64 \pi^2} 
\Biggl[ \ln\Bigl[ \frac{(1 \!-\! \epsilon)^2 H^2}{\mu_2^2}\Bigr] + 
\psi\Bigl(\frac12 \!+\! \nu\Bigr) + \psi\Bigl(\frac12 \!-\! \nu\Bigr)\Biggr]
\Biggr\} \Phi^3 \nonumber \\
& & \hspace{2.1cm} + \frac{\lambda}{96 \pi^2} \, (1 \!-\! 5\epsilon \!+\! 
3 \epsilon^2) H^2 \Phi + \frac{\lambda}2 \sum_{M=0}^{N} \delta i\Delta_M(x;x) 
\Phi + O(\hbar^2) \; . \qquad \label{hub1}
\end{eqnarray}
A result for this has already been reported in the literature for the
de Sitter case of $\epsilon =0$ and $\nu = \frac32$ \cite{Bilandzic:2007nb}.
Note that for $D=4$ the index (\ref{ournu}) becomes,
\begin{equation}
\nu^2 \Bigl\vert_{D=4} = \frac14 \Bigl(\frac{3 \!-\! \epsilon}{1 \!-\!
\epsilon}\Bigr)^2 - \frac{[\xi (12 \!-\! 6 \epsilon) \!+\! \frac12 \lambda 
\Phi_0^2]}{(1 - \epsilon)^2} = \frac14 - \frac{[(\xi \!-\! \frac16) R \!+\!
\frac12 \lambda \Phi^2]}{(1 \!-\! \epsilon)^2 H^2} \; .
\end{equation}
We will assume $\nu$ is positive (corresponding to inflation with small
$\Phi_0$) but one should bear in mind that it might be negative or even 
imaginary.

It remains to evaluate the lower limit contributions $\delta i\Delta_{M}(x;x)$
which were defined in equations (\ref{idD}-\ref{akl}). At coincidence we have
$\Vert \vec{x} - \vec{x}'\Vert = 0$ and $\eta = \eta'$. Only the $k=0$ term
of the sum in expression (\ref{idD}) contributes, and the sum over $\ell$
can be performed to give,
\begin{equation}
\delta i\Delta_{M}(x;x) = \frac{(1 \!-\! \epsilon)^2 H^2}{32 \pi^2} 
\, \frac{(4 k_0^2 \eta^2)^{M + \frac32 - \nu}}{\nu \!-\! M \!-\! \frac32}
\, \frac{\Gamma(2\nu \!-\! M) \Gamma(2\nu \!-\! 2 M)}{\Gamma(M \!+\! 1) 
\Gamma^2(\nu \!+\! \frac12 \!-\! M)} \; .
\end{equation}
It might be worth noting from relation (\ref{useful}) that the factor $k_0^2
\eta^2$ can be written in terms of the Hubble parameter and scale factor,
\begin{equation}
k_0^2 \eta^2 = \frac{k_0^2}{(1 \!-\! \epsilon)^2 H^2 a^2} \; . \label{krel}
\end{equation}
Note also that if the sum over $M$ is extended to infinity the result can be
expressed as a hypergeometric function,
\begin{eqnarray}
\lefteqn{\sum_{M=0}^{\infty} \delta i\Delta_M(x;x) = \frac{(1 \!-\! \epsilon)^2
H^2}{2 \pi^3} \, \frac{(\frac14 k_0^2 \eta^2)^{\frac32 - \nu}}{\nu \!-\! 
\frac32} } \nonumber \\
& & \hspace{2.5cm} \times 
\Gamma^2(\nu) \, \mbox{}_2F_3\Bigl( \frac12 \!-\! \nu, \frac32 \!-\! \nu;
1 \!-\!2\nu, 1 \!-\! \nu, \frac52 \!-\! \nu; -k_0^2 \eta^2 \Bigr) \; . \qquad
\end{eqnarray}

When the index $\nu$ approaches $\frac32 + N$, for some nonnegative integer
$N$ it is desirable to extract the infrared divergence from $\psi(\frac12 
- \nu)$ in expression (\ref{hub1}),
\begin{equation}
\psi\Bigl(\frac12 \!-\! \nu\Bigr) = \psi\Bigl(N \!+\! \frac52 \!-\! \nu\Bigr)
+ \sum_{M=-1}^N \frac1{\nu \!-\! M \!-\! \frac32} \; . \label{extract}
\end{equation}
A manifestly finite form can be obtained by combining the series in 
(\ref{extract}) with the series of lower limit contributions. This is
facilitated by the identity,
\begin{equation}
\Bigl(\xi \!-\! \frac16\Bigr) R + \frac{\lambda}2 \, \Phi^2 = -(1 \!-\!
\epsilon)^2 H^2 \Bigl(\nu^2 \!-\! \frac14\Bigr) \; .
\end{equation}
The final result for $V_{\rm hub}'(\Phi)$ is,
\begin{eqnarray}
\lefteqn{V_{\rm hub}'(\Phi) = \frac{\lambda}{96 \pi^2} \, 
(1 \!-\! 5\epsilon \!+\! 3 \epsilon^2) H^2 \Phi } \nonumber \\
& & + \Biggl\{ \xi + \frac{\lambda (\xi \!-\! \frac16)}{32 \pi^2} 
\Biggl[\ln\Bigl[ \frac{(1 \!-\! \epsilon)^2 H^2}{\mu_1^2}\Bigr] + 
\psi\Bigl(\frac12 \!+\! \nu\Bigr) + \psi\Bigl(N \!+\! \frac52 \!-\! \nu\Bigr)
\Biggr]\Biggr\} R \Phi \nonumber \\
& & + \Biggl\{ \frac{\lambda}6 + \frac{\lambda^2}{64 \pi^2} 
\Biggl[ \ln\Bigl[ \frac{(1 \!-\! \epsilon)^2 H^2}{\mu_2^2}\Bigr] + 
\psi\Bigl(\frac12 \!+\! \nu\Bigr) + \psi\Bigl(N \!+\! \frac52 \!-\! \nu\Bigr)
\Biggr] \Biggr\} \Phi^3 \nonumber \\
& & + \frac{(1\!-\!\epsilon)^2 H^2}{32 \pi^2} \sum_{M=-1}^{N} 
\frac1{\nu \!-\! M \!-\! \frac32} \, \Biggl\{ \frac{\Gamma(2\nu \!-\! M) 
\Gamma(2\nu \!-\! 2M) \, (4 k_0^2 \eta^2)^{M +\frac32 - \nu}}{2 
\Gamma(M \!+\! 1) \Gamma^2(\nu \!+\! \frac12 \!-\! M)} \nonumber \\
& & \hspace{7.5cm} -\Bigl(\nu^2 \!-\! \frac14\Bigr)\Biggr\} \, \lambda \Phi
+ O(\hbar^2) \; . \qquad \label{hub2}
\end{eqnarray}

\section{Field Strength, Time and Deceleration}

In the last section we obtained explicit forms (\ref{hub1}) and (\ref{hub2})
for $V_{\rm hub}'(\Phi)$. The purpose of this section is to discuss how the
Hubble Effective Potential depends upon the scalar field strength, upon time,
and upon the arbitrary constant deceleration. We begin by deriving a large
field expansion and checking the flat space correspondence limit against
the classic result of Coleman and Weinberg \cite{SCEW}. This suggests
symmetry breaking, which we confirm, for early times, by expanding around 
$\Phi = 0$. However, the explicit time dependence of the Hubble Effective 
Potential introduces a profound change: {\it prolonged time evolution during
inflation restores the $\Phi \rightarrow -\Phi$ symmetry.} We develop an
expansion for intermediate field strengths which shows that the symmetry
breaking minima come before this point. Then we investigate the explicit 
time dependence more fully, and the section closes with a discussion of the 
dependence upon $\epsilon$. In all cases we assume $0 \leq \epsilon < 1$, 
corresponding to inflation. We shall also equate the two mass scales $\mu_1 
= \mu_2 \equiv \mu$ and omit the $+ O(\hbar^2)$ at the end of each result 
for $V_{\rm hub}'(\Phi)$ in order to obtain more compact expressions.

When the scalar field strength is large the index $\nu$ becomes imaginary,
\begin{equation}
\nu = i \sqrt{ \frac{[(\xi \!-\! \frac16) R \!+\! \frac12 \lambda \Phi^2]}{
(1 \!-\! \epsilon)^2 H^2} - \frac14} \; .
\end{equation}
In this case it is best to expand the first form (\ref{hub1}). The key
relation we need comes from differentiating the logarithm of Stirling's
formula for the Gamma function,
\begin{equation}
\psi(z) = \ln(z) - \frac1{2 z} - \frac1{12 z^2} + \frac1{120 z^4} +
O\Bigl(\frac1{z^6}\Bigr) \; .
\end{equation}
The resulting expansion of the digamma function terms is,
\begin{eqnarray}
\lefteqn{\psi\Bigl(\frac12 \!+\! \nu\Bigr) + \psi\Bigl(\frac12 \!-\! \nu\Bigr)
= \ln\Biggl[ \frac{(\xi \!-\! \frac16) R \!+\! \frac12 \lambda \Phi^2}{
(1 \!-\! \epsilon)^2 H^2}\Biggr] } \nonumber \\
& & \hspace{1.5cm} -\frac13 \Biggl[ \frac{(1 \!-\! \epsilon)^2 H^2}{(\xi \!-\!
\frac16) R \!+\! \frac12 \lambda \Phi^2}\Biggr] -\frac1{15} \Biggl[ 
\frac{(1 \!-\! \epsilon)^2 H^2}{(\xi \!-\! \frac16) R \!+\! \frac12 \lambda 
\Phi^2}\Biggr]^2 + O(\Phi^{-6}) \; . \qquad
\end{eqnarray}
When the index is imaginary there are no infrared divergences at all and
the lower limit correction terms can be dropped. From (\ref{hub1}) we
see that the resulting large $\Phi$ form is,
\begin{eqnarray}
\lefteqn{V_{\rm hub}'(\Phi) = \xi R \Phi + \frac{\lambda}6 \, \Phi^3
+ \frac{\lambda}{96 \pi^2} \, (1 \!-\! 5\epsilon \!+\! 3 \epsilon^2) H^2 \Phi}
\nonumber \\
& & \hspace{1cm} + \frac{\lambda}{32 \pi^2} \Bigl[\Bigl(\xi \!-\! \frac16\Bigr)
R \!+\! \frac{\lambda}2 \Phi^2\Bigr] \Biggl\{ \ln\Bigl[\frac{(\xi \!-\! 
\frac16) R \!+\! \frac12 \lambda \Phi^2}{\mu^2}\Bigr] \nonumber \\
& & \hspace{1cm} -\frac13 \Bigl[ \frac{(1 \!-\! \epsilon)^2 H^2}{(\xi \!-\!
\frac16) R \!+\! \frac12 \lambda \Phi^2}\Bigr] -\frac1{15} \Bigl[ 
\frac{(1 \!-\! \epsilon)^2 H^2}{(\xi \!-\! \frac16) R \!+\! \frac12 \lambda 
\Phi^2}\Bigr]^2 + O(\Phi^{-6}) \Biggr\} \Phi \; . \qquad \label{large}
\end{eqnarray}
The coefficient of the logarithm term agrees, as it must, with equation (3.10) 
of the classic paper by Coleman and Weinberg \cite{SCEW}. It agrees as well
with the result previously obtained in de Sitter \cite{Bilandzic:2007nb} and
with what has been shown in general backgrounds \cite{BO}. This large field 
strength regime is also appropriate to a Higgs with $\xi \gg 1$, which has 
been proposed as a possible candidate for the inflaton \cite{BS,BKS}.

Expression (\ref{large}) suggests that symmetry breaking may occur, however,
care must be taken of the fact that (\ref{large}) is only the large field
form of a function that may not actually pass through zero for $\Phi \neq 0$.
Symmetry breaking in the large field form (\ref{large}) requires the logarithm 
to be negative, which is realized for,
\begin{equation}
\Bigl(\xi \!-\! \frac16\Bigr) R + \frac{\lambda}2 \, \Phi^2 < \mu^2 \; .
\label{con1}
\end{equation}
On the other hand, the large field expansion is valid for,
\begin{equation}
\Bigl(\xi \!-\! \frac16\Bigr) R + \frac{\lambda}2 \, \Phi^2 > 
(1 \!-\! \epsilon)^2 H^2 \; . \label{con2}
\end{equation}
These two conditions can only be consistent if,
\begin{equation}
(1 \!-\! \epsilon)^2 H^2 < \mu^2 \; . \label{con3}
\end{equation}
This is very dubious if the theory is renormalized at some low energy 
scale.\footnote{
Of course one could use the curved space renormalization group \cite{Wald}
to move the scale $\mu$ up to the level of primordial inflation. We could not
set $\mu$ to the instantaneous Hubble parameter $H(t)$ because that is
time dependent \cite{RGF}, but it would be perfectly possible to evolve
$\mu$ up to the constant $H_0$. Then the large logarithms $\ln(H^2/\mu^2)$ 
that appear in our results would be absorbed into redefinitions of the
conformal and quartic couplings, $\xi(\mu)$ and $\lambda(\mu)$. As long as 
the low scale $\lambda$ is small enough to dominate the large logarithms the
net effect would be no change, so we shall continue to imagine that $\mu$ is
a Standard Model scale very much smaller than the Hubble parameter. One
important consequence is that, when we assume ``minimal coupling,'' we mean 
$\xi(\mu) = 0$ {\it at Standard Model scales}.}

To check if symmetry breaking really occurs we need to expand 
$V_{\rm hub}'(\Phi)$ around $\Phi = 0$, and the best form for this is 
expression (\ref{hub2}). For small coupling ($\lambda \ll 1$) we can see
right away that the sign of the conformal coupling controls whether or
not the potential dips below zero,
\begin{equation}
V_{\rm hub}'(\Phi) = \xi R \Phi + O(\lambda) \; .
\end{equation}
For $\xi R < 0$ there will be symmetry breaking, whereas there will be
no symmetry breaking (at least near $\Phi = 0$) if $\xi R > 0$. The only
case requiring detailed analysis is minimal coupling ($\xi =0$), or near
minimal coupling.

When $\xi = 0$, the value of the index $\nu$ at $\Phi =0$ is,
\begin{equation}
\nu_0 \equiv \lim_{\xi, \Phi = 0} \nu = \frac12 \Bigl(\frac{3 \!-\! 
\epsilon}{1 \!-\! \epsilon}\Bigr) \; .
\end{equation}
We shall assume $0 < \epsilon < 1$, which implies $\nu_0 > \frac32$.
Let us also assume that $\nu_0$ lies in the range,
\begin{equation}
N + 1 < \nu_0 < N + 2 \; , \label{range}
\end{equation}
for some nonnegative integer $N$. Whether or not the potential falls below 
zero near $\Phi = 0$ is controlled by the second derivative,
\begin{eqnarray}
\lefteqn{\lim_{\xi = 0} V_{\rm hub}''(0) = \frac{\lambda H^2}{32 \pi^2} }
\nonumber \\
& & \hspace{-.4cm} \times \Biggl\{ -(2\!-\!\epsilon) \Biggl[
\ln\Bigl[ \frac{(1 \!-\! \epsilon)^2 H^2}{\mu^2}\Bigr] + \psi\Bigl(\frac12 
\!+\! \nu_0\Bigr) + \psi\Bigl(N \!+\! \frac52 \!-\! \nu_0\Bigr) \Biggr] 
-\frac{(5 \!-\! 4\epsilon)}3 \nonumber \\
& & \hspace{.4cm} - \sum_{M=0}^{N} \frac{(1 \!-\! \epsilon)^2
}{\nu_0 \!-\! M \!-\! \frac32} \Biggl[ \nu_0^2 \!-\! \frac14 - 
\frac{\Gamma(2\nu_0 \!-\! M) \Gamma(2\nu_0 \!-\! 2M)}{2 M! \Gamma^2(\nu_0
\!+\! \frac12 \!-\! M) (4 k_0^2 \eta^2)^{\nu_0 - M -\frac32 }}\Biggr]
\Biggr\} \; . \qquad \label{V''}
\end{eqnarray}

\begin{figure}
\begin{center}
\includegraphics[width=2.5in]{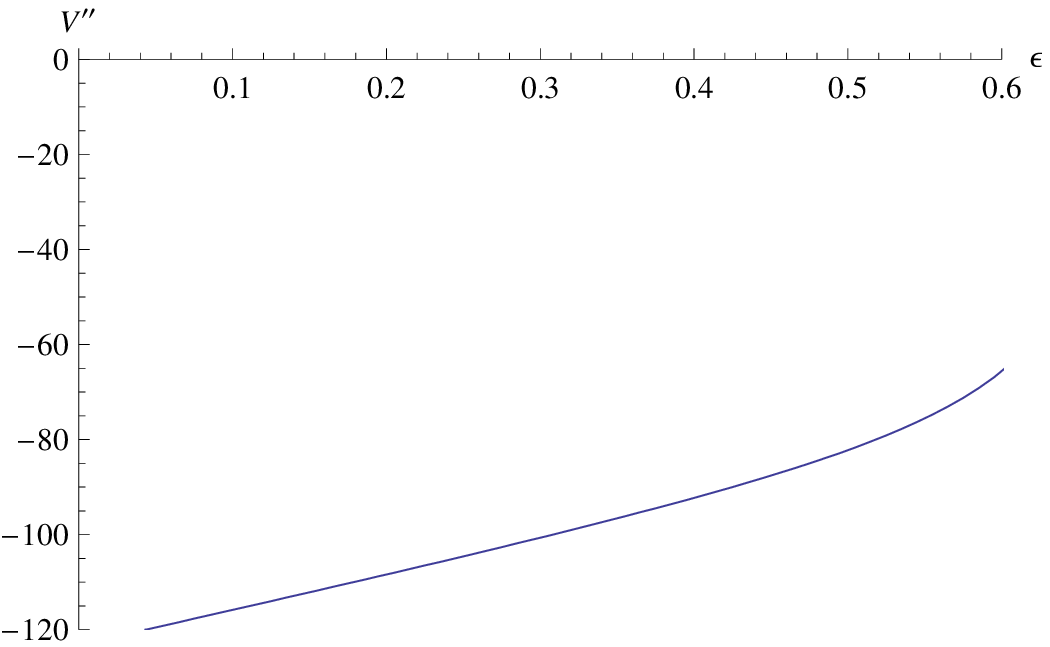}
\includegraphics[width=2.5in]{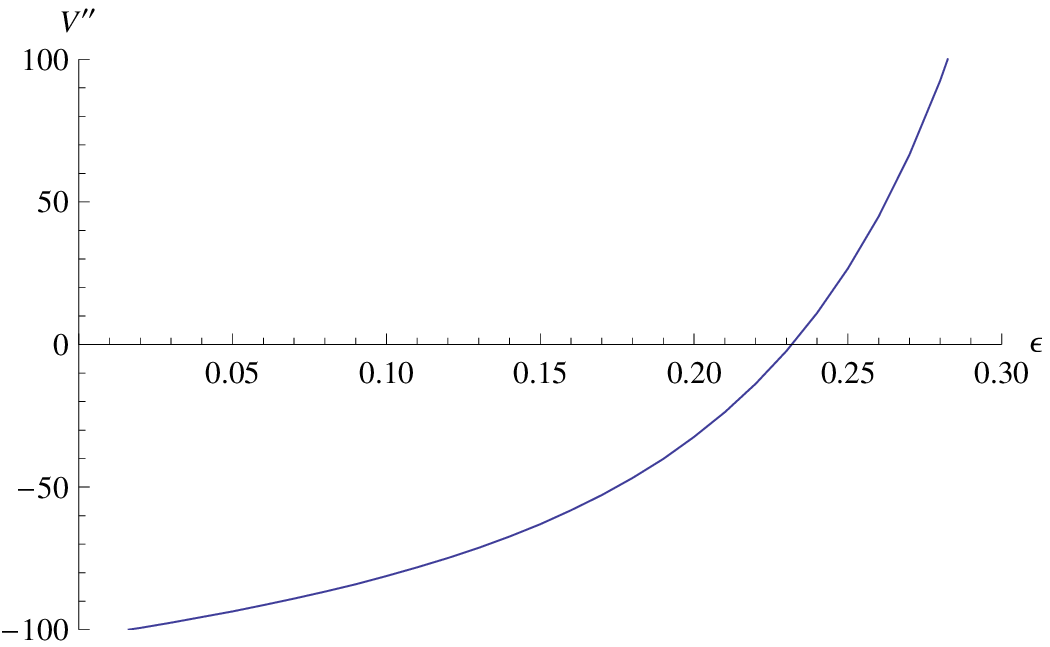}
\end{center}
\caption{Both plots show $V_{\rm hub}''(0) \div \lambda H^2/32\pi^2$ as a 
function of $\epsilon$ for $\xi = 0$ with $\mu/H_0 = 10^{-13}$ and $k_0/H_0
= 1$. The left plot is for $H_0 \eta = -1$ (shortly after the initial time)
and the right plot is for the later time $H_0 \eta = -10^{-2}$. Note that 
all values of $\epsilon$ within the plotted range show symmetry breaking
initially, and that time evolution tends to restore the symmetry. Symmetry
restoration takes place sooner for larger $\epsilon$.}
\label{V''plot}
\end{figure}

\begin{figure}
\begin{center}
\includegraphics[width=2.5in]{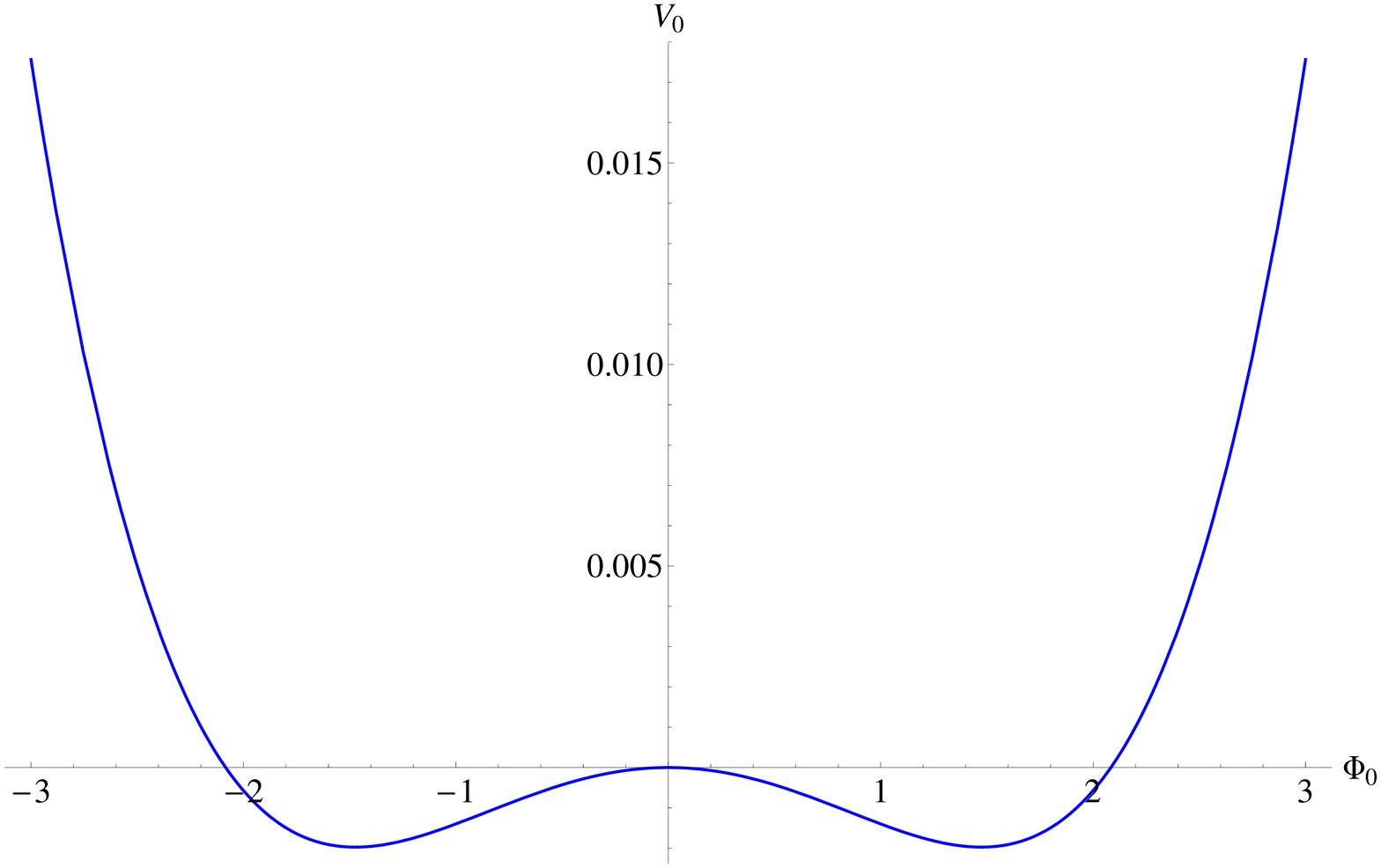}
\includegraphics[width=2.5in]{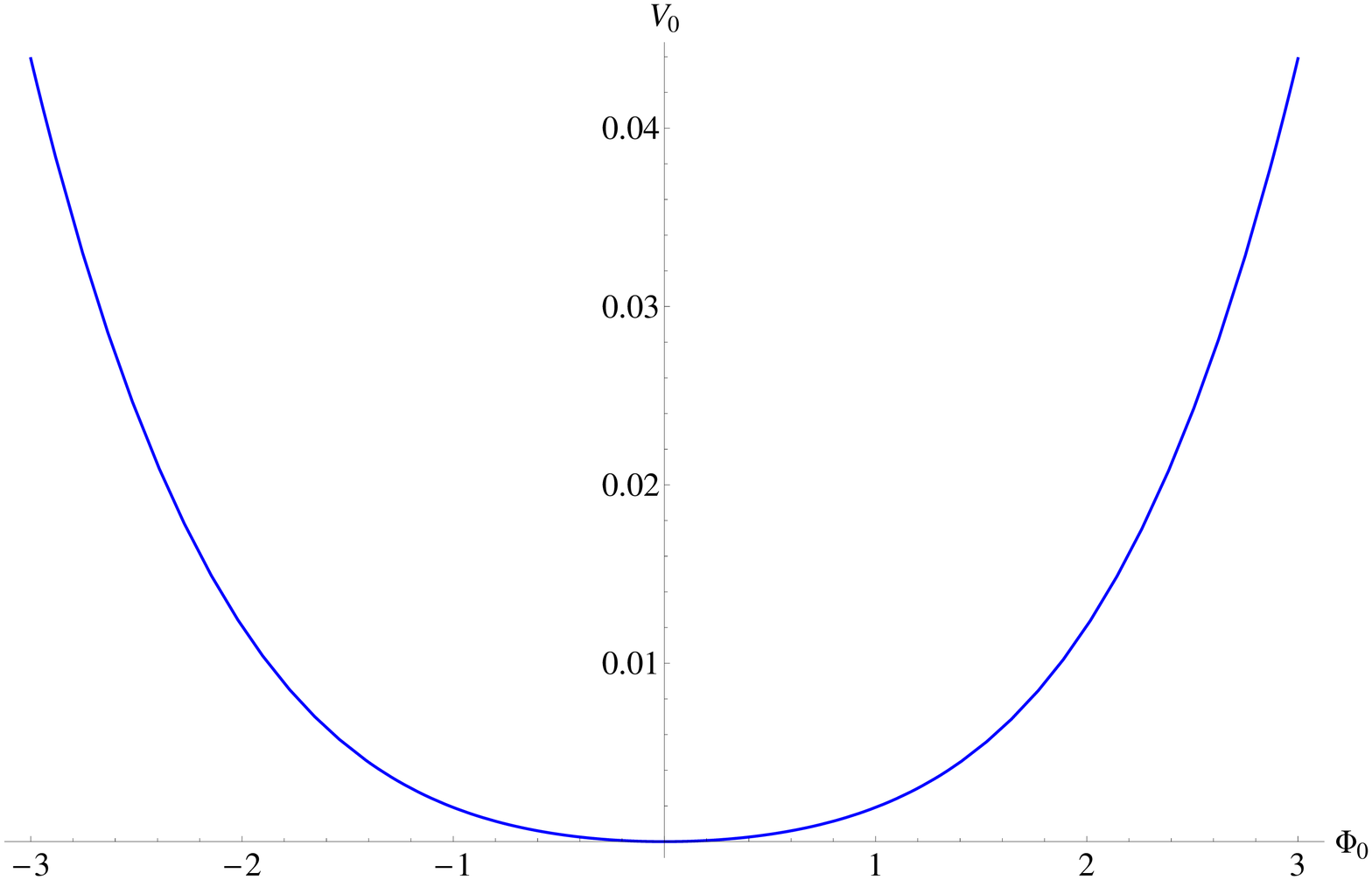}
\end{center}
\caption{Plot of the $V_0 \equiv V_{\rm hub}/H^4$ as a function of $\Phi_0$ 
for $\xi =0$, $\lambda = .01$ with $\epsilon = .1$, $\mu/H_0 = 10^{-13}$
and $k_0/H_0 = 1$. The left plot is for $H_0 \eta = -1$ (shortly after the
initial time) whereas the right plot is for $H_0 \eta = -10^{-5}$,
corresponding to about 12.8 e-foldings of further evolution. Symmetry
restoration occurs at about $H_0 \eta \approx -1.257 \times 10^{-4}$ for this
value of $\epsilon$.}
\label{earlylate}
\end{figure}

The sign of expression (\ref{V''}) is basically determined by the 
competition between its first and last terms,
\begin{equation}
\lim_{\xi = 0} V_{\rm hub}''(0) \approx \frac{\lambda H^2}{32 \pi^2} \Biggl\{
-(2\!-\!\epsilon) \ln\Bigl[ \frac{(1 \!-\! \epsilon)^2 H^2}{\mu^2}\Bigr] 
+ \frac{(1 \!-\! \epsilon)^2}{\nu_0 \!-\! \frac32} \, \frac{8 
\Gamma^2(\nu_0)}{\pi} \, \Bigl(\frac4{k_0^2 \eta^2}\Bigr)^{\nu_0 - \frac32}
\Biggr\} \; . \label{twoterms}
\end{equation}
The first (logarithm) term tends to make $V_{\rm hub}''(0) < 0$, whereas
the second ($\nu_0 -\frac32$ pole) term tends to make it positive. For a 
renormalization scale $\mu$ which is appropriate to the Standard Model, and 
a much higher scale Hubble parameter, the first term is large. For an infrared 
cutoff of $k_0 = H_0$, corresponding to no super-horizon modes in the initial 
state, the product $k_0^2 \eta^2$ is initially of order one, but rapidly 
approaches zero. This means that the second term is initially smaller in 
magnitude than the first, but grows without bound at late times. Hence we 
expect symmetry breaking shortly after the initial time of $\eta_0 = -1/(1
\!-\!\epsilon)H_0$, with symmetry restoration at late times. Because the 
exponent $\nu_0 -\frac32$ grows with $\epsilon$, we also expect the onset of 
symmetry restoration to come sooner for larger values of $\epsilon$. Both 
expectations are born out by explicit numerical evaluation of the full 
expression (\ref{V''}) --- not just the two terms of (\ref{twoterms}) --- as 
displayed in Fig.~\ref{V''plot}. The symmetry restoration effect can also
be seen by evolving the full potential for a fixed value of $\epsilon$, as
is displayed in Fig.~\ref{earlylate}.

We have seen how to obtain expansions for the large field and small
field regimes. As the magnitude of $\Phi$ increases, the value of $\nu$ 
decreases to zero, and then becomes imaginary. It is possible to develop
an expansion for the intermediate regime of $\Phi$, in which $\nu$ is
close to $\frac12$. The small quantity in this case is,
\begin{equation}
\frac{[(\xi \!-\! \frac16) R \!+\! \frac12 \lambda \Phi^2]}{(1 \!-\!
\epsilon)^2 H^2} \; . \label{smallpar}
\end{equation}
Regarding (\ref{smallpar}) as small amounts to expanding in powers of the
parameter $\Delta \nu$ defined as,
\begin{eqnarray}
\lefteqn{\Delta \nu \equiv \frac12 - \nu = \Biggl[ \frac{(\xi \!-\! \frac16)
R \!+\! \frac12 \lambda \Phi^2}{(1 \!-\! \epsilon)^2 H^2}\Biggr] } \nonumber \\
& & \hspace{3cm} + \Biggl[ \frac{(\xi \!-\! \frac16) R \!+\! \frac12 \lambda 
\Phi^2}{(1 \!-\! \epsilon)^2 H^2}\Biggr]^2 + 2 \Biggl[ \frac{(\xi \!-\! 
\frac16) R \!+\! \frac12 \lambda \Phi^2}{(1 \!-\! \epsilon)^2 H^2}\Biggr]^3
+ \dots \qquad
\end{eqnarray}
The digamma functions have the following expansions,
\begin{eqnarray}
\psi\Bigl(\frac12 \!+\! \nu\Bigr) & = & -\gamma -\sum_{n=2}^{\infty} 
\zeta(n) (\Delta \nu)^{n-1} \; , \\
\psi\Bigl(\frac12 \!-\! \nu\Bigr) & = & -\frac1{\Delta \nu} -\gamma -
\sum_{n=2}^{\infty} \zeta(n) (-\Delta \nu)^{n-1} \; , \qquad
\end{eqnarray}
where $\zeta(z)$ is the Riemann zeta function. It is also useful to note,
\begin{equation}
\Bigl(\xi \!-\! \frac16\Bigr) R + \frac{\lambda}2 \, \Phi^2 = (1\!-\!
\epsilon)^2 H^2 \Delta \nu (1 \!-\! \Delta \nu) \; .
\end{equation}
Because $\nu$ is less than the $N=0$ pole at $\nu = \frac32$, we can
ignore the infrared corrections (which fall off and are also suppressed 
by a prefactor proportional to $\Delta \nu^2$) and the result is,
\begin{eqnarray}
\lefteqn{V_{\rm hub}'(\Phi) = \Phi \Biggl\{ \Bigl[\frac1{18} \!+\! \frac23
\xi \!-\! \frac{\lambda}{576 \pi^2} \Bigr] R + (1 \!-\! \epsilon)^2 H^2 
\Delta \nu \Bigl[\frac13 \!-\! \frac13 \Delta \nu \!+\! \frac{\lambda}{32 
\pi^2}\Bigr] } \nonumber \\
& & \hspace{-.5cm} +\frac{\lambda (1 \!-\! \epsilon)^2 H^2 \Delta \nu (1 
\!-\! \Delta \nu)}{32 \pi^2} \Biggl[\ln\Bigl[ \frac{(1 \!-\!\epsilon)^2 H^2}{
\mu^2}\Bigr] \!-\! 2 \gamma \!-\!\! 2 \sum_{n=1}^{\infty} \zeta(2n \!+\! 1) 
\Delta \nu^{2n} \Biggr] \Biggl\} . \qquad \label{small}
\end{eqnarray}
For $\Delta \nu =0$ the result has the same sign as $\Phi$ for $\xi = 0$ and 
$\lambda < 32 \pi^2$, so we are beyond the symmetry breaking minima (if any), 
although still not in the large field regime.

We turn now to a systematic investigation of explicit time dependence. This 
enters the Hubble Effective Potential in two ways:
\begin{enumerate}
\item{From the factors of $\ln[H(t)]$ that combine with the logarithms of the
dimensional regularization mass scale $\mu_1 = \mu_2 = \mu$ in expressions
(\ref{hub1}) and (\ref{hub2}); and}
\item{From the factors of $(k_0^2 \eta^2)^{M +\frac32 -\nu}$ that multiply 
the infrared corrections of expression (\ref{hub2}).}
\end{enumerate}
The first of these has been included in the large field expansion
(\ref{large}) and in the small $\Delta \nu$ expansion (\ref{small}). 
If the field $\Phi(x)$ plays a role in low energy physics then the scale
$\mu$ is presumably in the Standard Model, and hence much less than the
Hubble parameter $H(t)$ at all times during primordial inflation. The
Hubble parameter does not change much during inflation so we expect these
logarithms to be large and slowly decreasing, which reduces the coefficients 
of the $R \Phi$ and $\Phi^3$ contributions to $V_{\rm hub}'(\Phi)$. We have 
seen from expressions (\ref{large}) and (\ref{small}) that this time 
dependence has little effect on symmetry breaking.

The second source of temporal dependence behaves very differently. Recall
that the conformal time $\eta$ begins at $\eta_0 =-1/(1\!-\!\epsilon) H_0$
during inflation and approaches zero at asymptotically late times. This 
means that the factors of $(k_0^2 \eta^2)^{M + \frac32 - \nu}$ grow like
powers of the scale factor for any $M$ such that $M < \nu -\frac32$. The
fastest growth comes from $M=0$, which contributes the following term to
$V_{\rm hub}'(\Phi)$,
\begin{equation}
\frac{\lambda (1 \!-\! \epsilon)^2 H^2}{32\pi^2} \, \frac{8 \Gamma^2(\nu)}{\pi}
\, \frac{\Phi}{\nu - \frac32} \Bigl(\frac4{k_0^2 \eta^2}\Bigr)^{\nu -\frac32}
\; . \label{M=0}
\end{equation}
The initial factor of $H^2(t)$ actually makes (\ref{M=0}) fall with time,
but it still grows with respect the $H^2 \Phi$ terms that would otherwise 
drive symmetry breaking. Another point to note is the dependence on the 
constant $\Phi_0$. Because the index $\nu$ decreases as a function of 
$\Phi_0^2$, eventually reaching zero and becoming imaginary, we see that 
(\ref{M=0}) vanishes in the large field limit. The same is true for the 
contributions from any $M$.

\begin{figure}
\begin{center}
\includegraphics[width=5in]{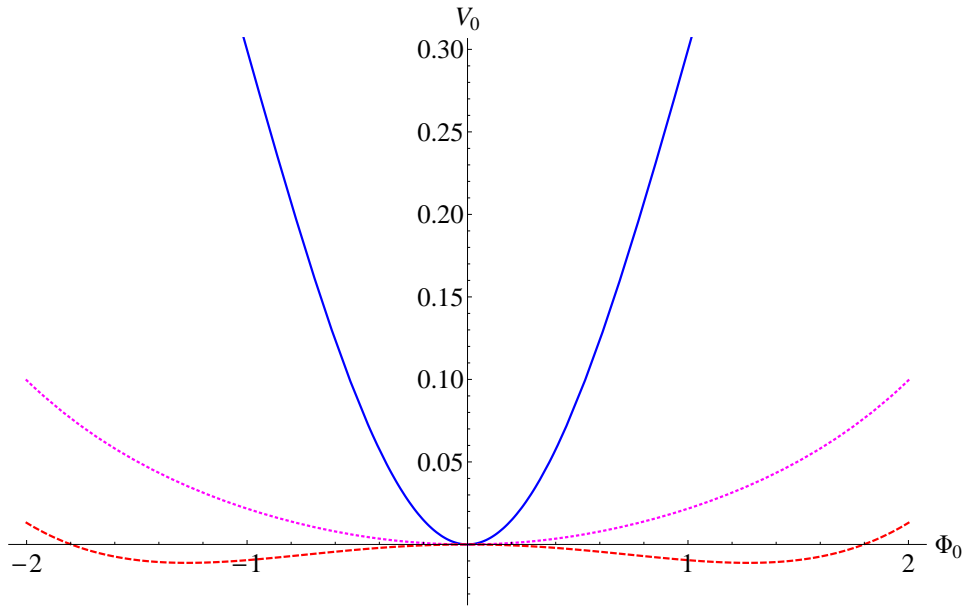}
\end{center}
\caption{Three different snapshots of $V_0 \equiv V_{\rm hub}/H^4$ as a 
function of $\Phi_0$ for $\xi =0$, $\lambda = .1$ with $\epsilon = .2$, 
$\mu/H_0 = 10^{-13}$ and $k_0/H_0 = 1$. The red dashed curve is for $H_0 
\eta = -10^{-1}$, the purple dotted curve is for $H_0 \eta = -10^{-3}$ 
(about 3.7 e-foldings later), and the blue solid curve is for $H_0 \eta = 
-10^{-5}$ (another 3.7 e-foldings later).}
\label{evolve}
\end{figure}

\begin{figure}
\begin{center}
\includegraphics[width=5in]{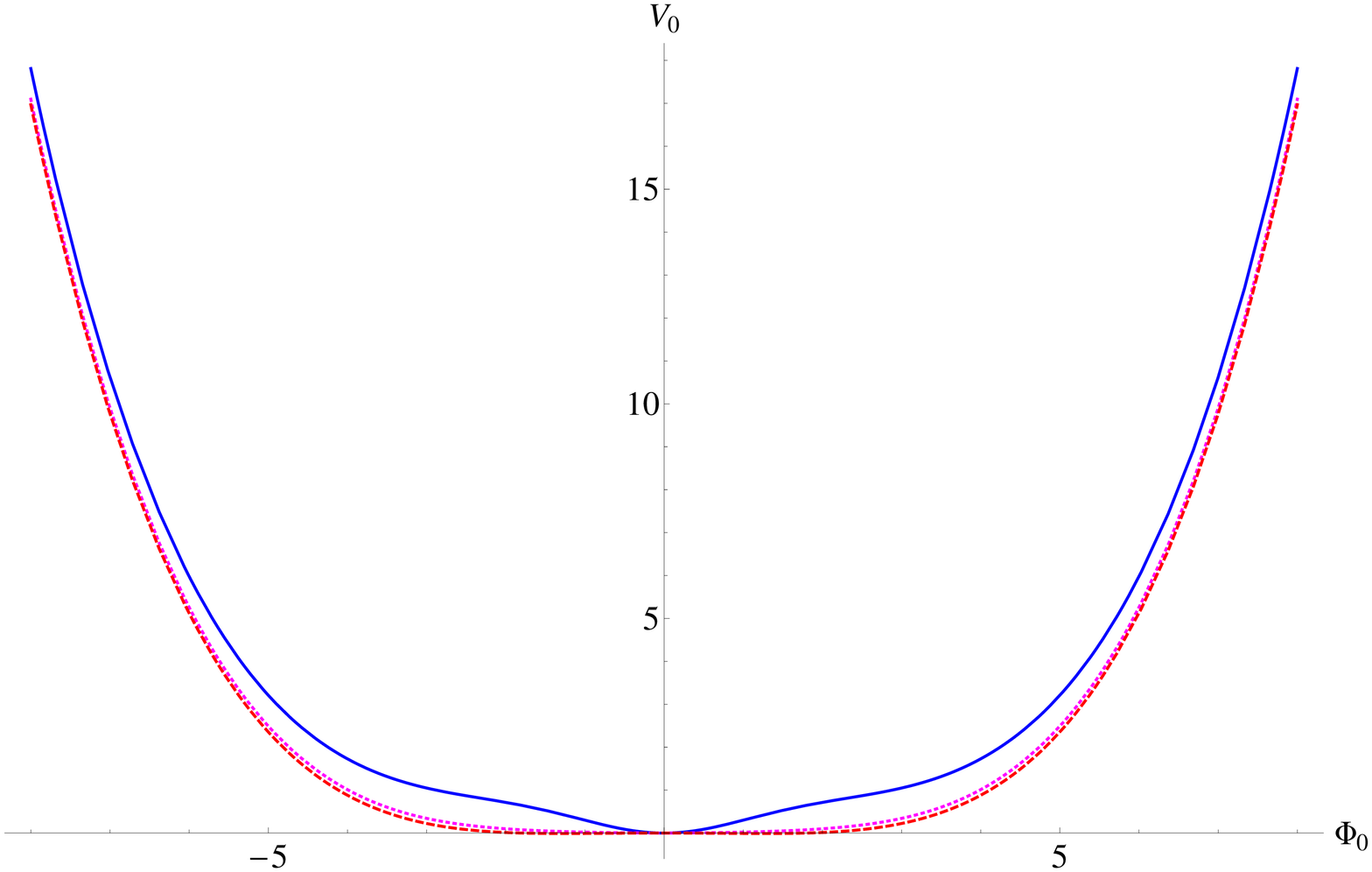}
\end{center}
\caption{The same three snapshots of $V_0 \equiv V_{\rm hub}/H^4$ as a 
function of $\Phi_0$ as in Fig.~\ref{evolve} over a larger range of $\Phi_0$.
For these parameters ($\lambda = .1$ and $\epsilon = .2$) the quantity 
$\Delta \nu \equiv \frac12 - \nu$ of the intermediate expansion (\ref{small}) 
vanishes at $\Phi_0 = \pm 6$. The index $\nu$ becomes imaginary for $\Phi_0 
= \pm \frac{14}{\sqrt{5}} \approx \pm 6.3$ and the large field expansion 
(\ref{large}) is valid by the end of the plotted range at $\Phi_0 = \pm 8$.}
\label{largescale}
\end{figure}

The various trends we have just described can be seen by plotting the
Hubble Effective Potential for the same parameters at different times.
To factor out the time dependence of the Hubble parameter we have scaled
the potential and the field by $H(t)$,
\begin{equation}
\Phi_0 \equiv \frac{\Phi(t)}{H(t)} \qquad , \qquad V_0(\Phi_0) \equiv 
\frac{V_{\rm hub}(\Phi)}{H^4(t)} \; .
\end{equation}
For the various parameters we chose,
\begin{equation}
\xi = 0 \quad , \quad \lambda = .1 \quad ,\quad \epsilon = \frac15 \quad , 
\quad \mu = 10^{-13} H_0 \quad , \quad k_0 = H_0 \; . \label{params}
\end{equation}
For these values, the transition from symmetry breaking to symmetry
restoration occurs at about $H_0 \eta \approx -5.05 \times 10^{-3}$.
Fig.~\ref{evolve} shows $V_0$ over the range $-2 < \Phi_0 < +2$ for
three different conformal times,
\begin{equation}
H_0 \eta = -10^{-1} \quad , \quad H_0 \eta = -10^{-3} \quad , \quad
H_0 \eta = -10^{-5} \; .
\end{equation}
The earliest time shows symmetry breaking, whereas the symmetry has been
restored by the second time, and the third plot shows that further time 
evolution makes the potential steeper at the origin. Fig.~\ref{largescale} 
shows the same three curves but over the expanded scale $-8 < \Phi_0 < +8$.
For parameters (\ref{params}) the center of the intermediate regime --- 
expansion (\ref{small}) --- comes at $\Phi_0 = \pm 6$, and the index $\nu$ 
goes imaginary at $\Phi_0 = \pm 14/\sqrt{5} \approx \pm 6.3$. The large 
field expansion (\ref{large}) is valid by $\Phi_0 = \pm 8$. Note that the 
steepening at late times is a small field effect; time evolution has
very little impact in the large field or even the intermediate field 
regime.

\begin{figure}
\begin{center}
\includegraphics[width=5in]{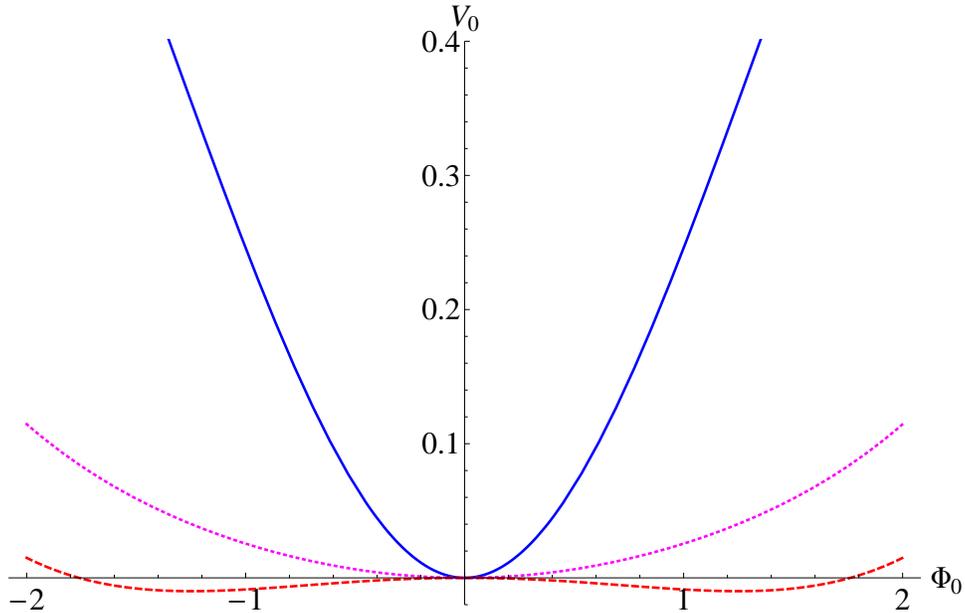}
\end{center}
\caption{$V_0 \equiv V_{\rm hub}/H^4$ as a function of $\Phi_0$ for three
different values of $\epsilon$. The other parameters are $\xi =0$, $\lambda = 
.1$, $\mu/H_0 = 10^{-13}$, $k_0/H_0 = 1$ and $H_0 \eta = -10^{-2}$. The red 
dashed curve is for $\epsilon = .1$, the purple dotted curve is for $\epsilon
= .3$, and the blue solid curve is for $\epsilon = .4$.}
\label{3epssmall}
\end{figure}

\begin{figure}
\begin{center}
\includegraphics[width=5in]{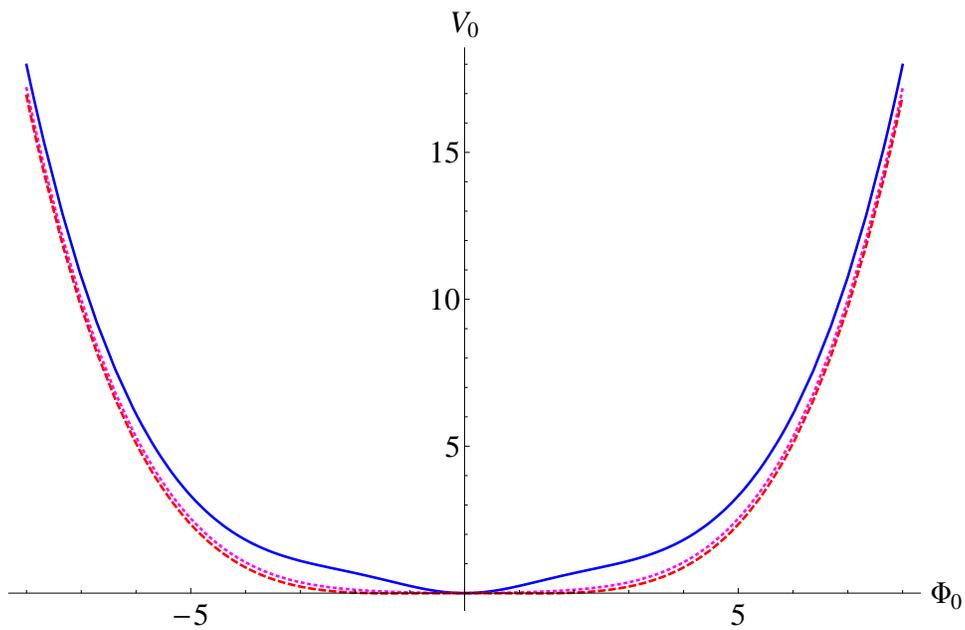}
\end{center}
\caption{The same plots of $V_0 \equiv V_{\rm hub}/H^4$ as a function of 
$\Phi_0$ as in Fig.~\ref{3epssmall} over a larger range of $\Phi_0$. For 
these parameters the center of the intermediate regime (\ref{small}) is
at about $\Phi_0 \approx \pm 6$, whereas the large field result (\ref{large})
applies by $\Phi_0 = \pm 8$.}
\label{3epslarge}
\end{figure}

We close with a comment on $\epsilon$ dependence. This is only very weak 
in the large field regime (\ref{large}), as required by the need to agree
with the result of Coleman and Weinberg \cite{SCEW}. On the other hand, 
$\epsilon$ dependence is an order one effect in the intermediate regime 
(\ref{small}) --- recall that $R = 6(2 -\epsilon) H^2$ and that $\Delta \nu$
vanishes for $\Phi_0^2 = 2(2\!-\!\epsilon)/\lambda$. In the small
field regime the value of $\epsilon$ controls the rate at which time 
evolution causes the $M = 0$ infrared contribution (\ref{M=0}) dominate. 
One can see all this by plotting $V_0$ versus $\Phi_0$ for different values 
of $\epsilon$ at the same time. In Fig.~\ref{3epssmall} and 
Fig.~\ref{3epslarge} we have chosen,
\begin{equation}
\xi = 0 \quad , \quad \lambda = .1 \quad ,\quad \mu = 10^{-13} H_0 \quad , 
\quad k_0 = H_0 \quad , \quad H_0 \eta = -10^{-2} \; ,
\end{equation}
for three different values of $\epsilon$,
\begin{equation}
\epsilon = .1 \quad , \quad \epsilon = .3 \quad , \quad \epsilon = .4 \; .
\end{equation}
In Fig.~\ref{3epssmall} the three curves are plotted over the range
$-2 < \Phi_0 < +2$. The plot for $\epsilon = .1$ shows symmetry breaking,
whereas the symmetry has been restored for $\epsilon = .3$, and the plot
for $\epsilon = .4$ shows that the potential steepens with increasing 
$\epsilon$. In Fig.~\ref{3epslarge} the same three curves are plotted over 
the expanded range $-8 < \Phi_0 < +8$. For these parameters the center
of the intermediate regime (\ref{small}) is at about $\Phi_0 \approx \pm 6$,
and the large field regime (\ref{large}) pertains by $\Phi_0 = \pm 8$.
One can see that the curve for $\epsilon = .4$ is still noticeably above
the other curves in the intermediate regime, but the three curves are
almost indistinguishable in the large field regime.

\section{Discussion}

We have introduced a new sort of effective potential which might be useful
in cosmological settings for which the curvature dominates over other 
dimensionful parameters. Instead of evaluating the effective action for a
constant background scalar, we evaluate it at a background that scales with
the Hubble parameter,
\begin{equation}
\Phi(t,\vec{x}) \longrightarrow \Phi_0 \times H(t) \; .
\end{equation}
In section 2 we show how to obtain the propagator (which is the limiting
step on any computation of the effective action) for a massless scalar
with arbitrary conformal coupling, in an FRW geometry of arbitrary but
constant deceleration. In section 3 we evaluated the Hubble Effective
Potential at one loop order for a quartic self-coupling.

Section 4 was spent discussing the dependence upon field strength, time
and deceleration under the assumption of primordial inflation far above
Standard Model scales. We obtained expansions valid for large field strength
(\ref{large}) and for intermediate field strength (\ref{small}), and we
made a careful study of the curvature (\ref{V''}) at $\Phi_0 = 0$.
Of course the large field limit agrees with the flat space result of
Coleman and Weinberg \cite{SCEW}, which suggests that symmetry breaking
may occur. We confirmed that it does occur, but only at early times. As 
time progresses the $\Phi \rightarrow -\Phi$ symmetry is restored, with
the effect coming sooner the larger the deceleration. Subsequent time
evolution steepens the potential near the origin, without of course affecting
its large field limit.

The fascinating time dependence of the Hubble Effective Potential seems to
mostly derive from inflationary particle production. (There is also a small
effect due to the time dependent Hubble parameter declining relative to the
fixed scale of renormalization.) One can understand what is happening on a
simple, qualitative level by noting that inflationary particle production
will, of necessity, increase the field strength. So it drives the field
away from the small symmetry breaking minima. This is why the effective
potential is driven up, at small field strength, but not affected much at
large field strength.

One consequence of the explicit time dependence we have found is that
the effective field $\Phi(t)$ cannot really evolve according to the 
Hubble-scaling ansatz (\ref{hubbak}) which made the calculation possible!
This does not mean what we have done is irrelevant, any more than it
would be irrelevant to use the effective potential to show that the field
cannot be constant. We now have a {\it proof} that the field cannot
exactly track $H(t)$. Further, as long as the time dependent terms are
small, the Hubble Effective Potential probably describes the actual 
evolution well, just as the effective potential is assumed to do for
slowly varying fields. At the very least, one can use our formalism to
determine {\it which way} the actual background is pushed away from 
exact Hubble-scaling. 

The effective potential we obtain is quite similar to those derived
on de Sitter background for this same model \cite{Bilandzic:2007nb},
for Yukawa theory \cite{MW} and for scalar quantum 
electrodynamics (SQED) \cite{SQED}. One striking feature of those models is
that the effective potential which describes scalar evolution is not quite
the same as the one that appears in the effective stress tensor and would
govern the gravitational response if back-reaction were included. This can
have interesting consequences. For example, although the effective potential 
for SQED is minimized at zero field strength, the vacuum energy actually 
decreases, for a while, as the field strength increases \cite{SQED}. The 
reason for the differences was traced to the fact that the effective potential
depends nontrivially upon the metric, so its variation --- which gives
the stress tensor --- must reflect this dependence \cite{MW,SQED}. The
same thing should be true for our model, and it would be quite interesting
to quantify this expectation by working out the one loop correction to the 
gravitational field equations. It would be particularly interesting to explore
the gravitational consequences of symmetry restoration.

Finally, we should contrast our view of the infrared issue with some
different approaches that have been suggested to dealing with infrared
divergences in loop corrections to the power spectrum \cite{mainstream}. 
One might argue that because super-horizon modes have no apparent spatial 
variation, they are not observable, and the results of quantum field 
theory must be changed to exclude them. It seems to us that spatial 
constants {\it are} observable --- indeed, the current universe seems to 
be experiencing the effect of something very much like a small cosmological 
constant \cite{Yun}! 

We also believe that the infrared divergences of loop corrections to the power 
spectrum are clear analogs of the old problem which afflicts the propagator 
of a massless, minimally coupled scalar on many cosmological backgrounds 
\cite{FP}. In that case the infrared divergence results, {\it as infrared
divergences always do in quantum field theory}, from posing an unphysical 
question. The physically dubious feature of the question is that one can 
prepare coherent Bunch-Davies vacuum over an infinite spatial section. 
Finite results are obtained when this assumption is relaxed, either by 
working on a compact space \cite{TW3} or by making the super-horizon modes 
of the initial state less singular than Bunch-Davies \cite{AV}. This sort 
of resolution leads to finite results which, however, grow with time as more
and more initially sub-horizon modes are redshifted to long wave lengths
\cite{JMP,JP2,OW,IRlogs}.

The fascinating secular effect of symmetry restoration we have found derives
from precisely this source, and it seems to have observable consequences.
For example, suppose we consider the scalar to be the Standard Model Higgs,
and suppose that it is effectively massless and minimally coupled during
primordial inflation. During the early stages of inflation, when the 
symmetry is broken, the fermions of the Standard Model will be massive, with 
the expected mass of the order $m_\psi\sim y H \Phi_0$, where $y$ is the 
Yukawa coupling and $\Phi_0\sim 1$. When $m_\psi\leq H$ (which is valid for
most Standard Model fermions with $y \ll 1$) they will be produced during 
inflation, reaching the occupation number \cite{Garbrecht:2006jm},
\begin{equation}
n_\psi \simeq \frac1{\exp(2\pi m_\psi/H)+1} \sim \frac1{\exp(2\pi y\Phi_0)+1}
\simeq 1/2 \; .
\end{equation}
This is a good estimate as long as $H$ changes adiabatically slowly in time,
i.e. when $\epsilon\ll 1$. (For a finite $\epsilon$ this formula will
probably be corrected to, $n_\psi = 1/[\exp(2\pi
m_\psi/[(1-\epsilon)H])+1]$.)  Now, if inflation is of a relatively
short duration, the symmetry may be restored quite late during
inflation, in which case some of the fermions may survive and carry an
observable information about the duration of inflation.

Another source of potentially observable effects is in cosmological 
perturbations. The effective potential we have been computing presumably 
affects gravity --- although probably not quite the way a classical scalar 
potential would \cite{MW,SQED}. That will change the geometry and alter the 
spectrum of density perturbations. And note that, whereas quantum 
gravitational corrections are necessarily suppressed by a factor of the 
small loop counting parameter $G H^2 \ltwid 10^{-12}$, the effects of
quantum {\it matter} loops need only be suppressed by Standard Model 
couplings.

\centerline{\bf Acknowledgements}

We are grateful for conversations on this subject with N. C. Tsamis.
This work was partially supported by FOM grant 07PR2522, by Utrecht
University, by European Union grant INTERREG-IIIA, by NSF grant
PHY-0653085, by the Project of Knowledge Innovation Program (PKIP) of
the Chinese Academy of Sciences (Grant No.~KJCX2.YW.W10) and by the 
Institute for Fundamental Theory at the University of Florida.


\begin{thebibliography}{99}

\bibitem{JS1} J. Schwinger, Phys. Rev. {\bf 82} (1951) 664.

\bibitem{BSD1} B. S. DeWitt, {\it The Dynamical Theory of Groups and Fields}
(Gordon and Breach, New York, 1965).

\bibitem{LFA} L. F. Abbott, Acta Phys. Polon. {\bf B13} (1982) 33.

\bibitem{JS2} J. Schwinger, J. Math. Phys. {\bf 2} (1961) 407.

\bibitem{others} K. T. Mahanthappa, Phys. Rev. {\bf 126} (1962) 329;
P. M. Bakshi and K. T. Mahanthappa, J. Math. Phys. {\bf 4} (1963) 1; J. 
Math. Phys. {\bf 4} (1963) 12;
L. V. Keldysh, Sov. Phys. JETP {\bf 20} (1965) 1018;
R. D. Jordan, Phys. Rev. {\bf D33} (1986) 444;
K. C. Chou, Z. B. Su, B. L. Hao and L. Yu, Phys. Rept. {\bf 118} (1985) 1;
E. Calzetta and B. L. Hu, Phys. Rev. {\bf D35} (1987) 495.

%\cite{Berges:2004yj}
\bibitem{Berges:2004yj}
  J.~Berges,
  %``Introduction to nonequilibrium quantum field theory,''
  AIP Conf.\ Proc.\  {\bf 739} (2005) 3
  [arXiv:hep-ph/0409233].
  %%CITATION = APCPC,739,3;%%

\bibitem{FW} L. H. Ford and R. P. Woodard, Class. Quant. Grav. {\bf 22} (2005)
1637, gr-qc/0411003.

\bibitem{BSD2} B. S. DeWitt and R. W. Brehme, Ann. Phys. {\bf 9} (1960) 220.

\bibitem{RTS} R. T. Seeley, Proc. Symp. Pure Math. {\bf 10} (1967) 288.

\bibitem{BSD3} B. S. DeWitt, Phys. Rept. {\bf 19} (1975) 295.

\bibitem{SMC} S. M. Cristensen, Phys. Rev. {\bf D17} (1978) 946.

\bibitem{BV} A. O. Barvinsky and G. A. Vilkovisky, Phys. Rept. {\bf 119}
(1985) 1.

\bibitem{us} T. M. Janssen, S. P. Miao, T. Prokopec and R. P. Woodard,
Class. Quant. Grav. {\bf 25} (2008) 245013, arXiv:0808.2449.

\bibitem{Bunch:1977sq}
  T.~S.~Bunch and P.~C.~W.~Davies,
  %``Covariant Point Splitting Regularization For A Scalar Quantum Field In A
  %Robertson-Walker Universe With Spatial Curvature,''
  Proc.\ Roy.\ Soc.\ Lond.\  A {\bf 357} (1977) 381.
  %%CITATION = PRSLA,A357,381;%%

\bibitem{JP1}
  T.~Janssen and T.~Prokopec,
  %``A graviton propagator for inflation,''
   Class. Quant. Grav. {\bf 25} (2008) 055007
  arXiv:0707.3919 [gr-qc].
  %%CITATION = ARXIV:0707.3919;%%

\bibitem{JMP}
  T.~Janssen, S.~P.~Miao and T.~Prokopec,
  %``Graviton one-loop effective action and inflationary dynamics,''
  arXiv:0807.0439 [gr-qc].
  %%CITATION = ARXIV:0807.0439;%%

\bibitem{FP}
  L.~H.~Ford and L.~Parker,
  %``Infrared Divergences In A Class Of Robertson-Walker Universes,''
  Phys.\ Rev.\  D {\bf 16} (1977) 245.
  %%CITATION = PHRVA,D16,245;%%

\bibitem{JP2}
  T.~Janssen and T.~Prokopec,
  %``Implications of the graviton one-loop effective action on the dynamics of
  %the Universe,''
  arXiv:0807.0447 [gr-qc].
  %%CITATION = ARXIV:0807.0447;%%

\bibitem{AV}
A.~Vilenkin,
  %``Quantum Fluctuations In The New Inflationary Universe,''
  Nucl.\ Phys.\  B {\bf 226}, 527 (1983).
  %%CITATION = NUPHA,B226,527;%%

\bibitem{Fulling:1978ht}
  S.~A.~Fulling, M.~Sweeny and R.~M.~Wald,
  %``Singularity Structure Of The Two Point Function In Quantum Field Theory In
  %Curved Space-Time,''
  Commun.\ Math.\ Phys.\  {\bf 63} (1978) 257.
  %%CITATION = CMPHA,63,257;%%

\bibitem{TW3}
  N.~C.~Tsamis and R.~P.~Woodard,
  %``The Physical basis for infrared divergences in inflationary quantum
  %gravity,''
  Class.\ Quant.\ Grav.\  {\bf 11}, 2969 (1994).
  %%CITATION = CQGRD,11,2969;%%

\bibitem{OW} V. K. Onemli and R. P. Woodard, Class. Quant. Grav. {\bf 19} 
(2002) 4607, gr-qc/0204065; Phys. Rev. {\bf D70} (2004) 107301, gr-qc/0406098.

\bibitem{ITTW} J. Iliopoulos, T. N. Tomaras, N. C. Tsamis and R. P. Woodard,
Nucl. Phys. {\bf B534} (1998) 419, gr-qc/9801028.

 %\cite{Bilandzic:2007nb}
\bibitem{Bilandzic:2007nb}
   A.~Bilandzic and T.~Prokopec,
   %``Quantum radiative corrections to slow-roll inflation,''
   Phys.\ Rev.\  D {\bf 76} (2007) 103507
   [arXiv:0704.1905 [astro-ph]].
   %%CITATION = PHRVA,D76,103507;%%

\bibitem{SCEW} S. R. Coleman and E. J. Weinberg, Phys. Rev. {\bf D7} (1973)
1888.

\bibitem{BO} I. L. Buchbinder and S. D. Odintsov, Class. Quantum Grav. {\bf 2}
(1985) 721.

\bibitem{BS} F. Bezrukov and M. Shaposhnikov, Phys. Lett. {\bf B659} (2008)
703, arXiv:0710.3755.

\bibitem{BKS} A. O. Barvinsky, A. Yu. Kamenshchik and A. A. Starobinsky,
JCAP {\bf 0811} (2008) 021, arXiv:0809.2104; 
A. O. Barvinsky, A. Yu. Kamenshchik, C. Kiefer, A. A. Starobinsky and C. 
Steinwachs, arXiv:0904.1698.

\bibitem{Wald} S. Hollands and R. M. Wald, Commun. Math. Phys. {\bf 237}
(2003) 123, gr-qc/0209029.

\bibitem{RGF} R. P. Woodard, Phys. Rev. Lett. {\bf 101} (2008) 081301, 
arXiv:0805.3089.

\bibitem{MW} S. P. Miao and R. P. Woodard, Phys. Rev. {\bf D74} (2006) 
044019, gr-qc/0602110.

\bibitem{SQED} T. Prokopec, N. C. Tsamis and R. P. Woodard, Annals Phys. 
{\bf 323} (2008) 1324, arXiv:0707.0847.

\bibitem{mainstream} D. H. Lyth, JCAP {\bf 0712} (2007) 016, arXiv:0707.0361;
N. Bartolo, S. Matarrese, M. Pietroni, A. Riotti and D. Seery, JCAP {\bf 0801}
(2008) 015, arXiv:0711.4263;
Y. Urakawa and T. Tanaka, ``No influence on observation from IR divergence 
during Inflation I,'' arXiv:0902.3209.

\bibitem{Yun} Y. Wang and P. Mukherjee, Astrophys. J. {\bf 650} (2006) 1,
astro-ph/0604051; U. Alam, V. Sahni and A. A. Starobinsky, JCAP {\bf 0702}
011, 2007, astro-ph/0612381.

\bibitem{IRlogs}  A. A. Starobinsky and J. Yokoyama, Phys. Rev. {\bf D50} 
(1994) 6357, astro-ph/9407016; 
N. C. Tsamis and R. P. Woodard, Nucl. Phys. {\bf B474} (1996) 235,
hep-ph/9602315; Phys. Rev. {\bf D54} (1996) 2621, hep-ph/9602317; 
Ann. Phys. {\bf 253} (1997) 1, hep-ph/9602316; arXiv:0807.5006;
T. Prokopec, O. Tornkvist and R. P. Woodard, Phys. Rev. Lett. {\bf 89} (2002) 
101301, astro-ph/0205331; Ann. Phys.  {\bf 303} (2003) 251, gr-qc/0205130;
T. Prokopec and R. P. Woodard, JHEP {\bf 0310} (2003) 059,
astro-ph/0309593; Ann. Phys. {\bf 312} (2004) 1, gr-qc/0310056;
T. Prokopec and E. Puchwein, JCAP {\bf 0404} (2004) 007, astro-ph/\-0312274;
T. Prokopec, N.C. Tsamis and R. P. Woodard, Class. Quant. Grav. {\bf 24} 
(2007) 201, gr-qc/0607094; Phys. Rev. {\bf D78} (2008) 043523, arXiv:0802.3673;
S. Weinberg, Phys. Rev. {\bf D72} (2005) 043514, hep-th/0506236;
Phys. Rev. {\bf D74} (2006) 023508, hep-th/0605244;
T. Brunier, V. K. Onemli and R. P. Woodard, Class. Quant. Grav.
{\bf 22} (2005) 59, gr-qc/0408080;
D. Boyanovsky, H. J. de Vega and N. G. Sanchez, Nucl. Phys. {\bf B747} (2006) 
25, astro-ph/0503669; Phys. Rev. {\bf D72} (2005) 103006, astro-ph/0507596;
M. Sloth, Nucl. Phys. {\bf B748} (2006) 149, astro-ph/0604488;
K. Chaicherdsakul, Phys. Rev. {\bf D75} (2007) 063522, hep-th/0611352;
M. van der Meulen and J. Smit, JCAP {\bf 0711} (2007) 023, arXiv:\-0707.0842;
E. O. Kahya and V. K. Onemli, Phys. Rev. {\bf D76} (2007)
043512, gr-qc/0612026;
S. P. Miao and R. P. Woodard, Class. Quant. Grav. {\bf 23} (2006) 1721, 
gr-qc/0511140; Phys. Rev. {\bf D74} (2006) 024021, gr-qc/0603135; 
Class. Quant. Grav. {\bf 25} (2008) 145009, arXiv:0803.2377;
Y. Urakawa and K. I Maeda, Phys. Rev. {\bf D78} (2008) 064004, arXiv:0801.0126;
A. Riotto and M. Sloth, JCAP {\bf 0804} (2008) 030, arXiv:0801.1845;
K. Enqvist, S. Nurmi, D. Podolsky and G. I. Rigopoulos, JCAP {\bf 0804} (2008) 
025, arXiv:0802.0395; 
K. Enqvist, S. Nurmi and G. I. Rigopoulos, JCAP {\bf 0810} (2008) 013,
arXiv:0810.0382; 
A. E. Romano and M. Sasaki, Phys. Rev. {\bf D78} (2008) 103522, 
arXiv:0809.5141.

 %\cite{Garbrecht:2006jm}
\bibitem{Garbrecht:2006jm}
  B.~Garbrecht and T.~Prokopec,
  %``Fermion mass generation in de Sitter space,''
  Phys.\ Rev.\  D {\bf 73} (2006) 064036
  [arXiv:gr-qc/0602011].
  %%CITATION = PHRVA,D73,064036;%%

\end{thebibliography}
\end{document}